\documentclass[a4paper,11pt]{article}
\usepackage{jcappub} 
\usepackage{lineno}

\usepackage{dcolumn}
\usepackage{verbatim}

\usepackage{breakurl}

\usepackage{lmodern}

\pdfoutput=1
\usepackage{graphicx}
\usepackage{bm}
\usepackage{amsmath}
\usepackage{amsfonts}
\usepackage{amssymb}
\usepackage{multirow}
\usepackage[capitalise]{cleveref}
\usepackage{acro}
\usepackage[utf8]{inputenc}

\usepackage{tikz}
\usepackage{pgfplots}
\usepackage{subcaption}

\crefname{figure}{Figure}{Figures}

\newcommand{\be}{{\begin{eqnarray}}}
\newcommand{\ee}{{\end{eqnarray}}}

\newcommand{\overbar}[1]{\mkern 1.5mu\overline{\mkern-1.5mu#1\mkern-1.5mu}\mkern 1.5mu}

\newcommand{\fnl}{f_\mathrm{NL}}

\newcommand{\gnl}{g_\mathrm{NL}}

\newcommand{\cH}{\mathcal{H}}

\newcommand{\cO}{\mathcal{O}}

\newcommand{\bk}{\mathbf{k}}
\newcommand{\bq}{\mathbf{q}}
\newcommand{\bp}{\mathbf{p}}
\newcommand{\bl}{\mathbf{l}}
\newcommand{\bx}{\mathbf{x}}

\newcommand{\bn}{\mathbf{n}}

\newcommand{\ud}{\mathrm{d}}
\newcommand{\uin}{\mathrm{in}}

\newcommand{\uGW}{\mathrm{gw}}
\newcommand{\ung}{\mathrm{ng}}
\newcommand{\uc}{\mathrm{c}}
\newcommand{\Beq}{\begin{align}}
\newcommand{\Eeq}{\end{align}}
\newcommand{\Glm}{\mathcal{G}_{\ell_1 \ell_2 \ell_3}^{m_1 m_2 m_3}}

\DeclareAcronym{GW}{
  short = GW,
  long = gravitational wave ,
  short-plural = s ,
}
\DeclareAcronym{LIGO}{
  short =LIGO ,
  long = Laser Interferometer Gravitational-Wave Observatory ,
  short-plural = ,
}
\DeclareAcronym{LVK}{
  short = LVK ,
  long = {LIGO, Virgo, and KAGRA},
  short-plural = ,
}

\DeclareAcronym{SGWB}{
  short = SGWB ,
  long = stochastic gravitational-wave background ,
  short-plural = s ,
}
\DeclareAcronym{GWB}{
  short = GWB ,
  long = gravitational-wave background ,
  short-plural = s ,
}
\DeclareAcronym{CBC}{
  short = CBC ,
  long = compact binary coalescence ,
  short-plural = s ,
}
\DeclareAcronym{BH}{
  short = BH ,
  long = black hole ,
  short-plural = s ,
}
\DeclareAcronym{BBH}{
  short = BBH ,
  long = binary black hole ,
  short-plural = s ,
}
\DeclareAcronym{PBH}{
  short = PBH ,
  long = primordial black hole ,
  short-plural = s ,
}
\DeclareAcronym{SNR}{
  short = SNR ,
  long = signal-to-noise ratio ,
  short-plural = s ,
}
\DeclareAcronym{IMRPPv2}{
  short = ,
  long = {\normalsize IMRP}{\footnotesize HENOM}{\normalsize P}v2 ,
  short-plural = ,
}
\DeclareAcronym{PTA}{
  short = PTA ,
  long = pulsar timing array ,
  short-plural = s ,
}
\DeclareAcronym{SFR}{
  short = SFR ,
  long = star formation rate ,
  short-plural =  ,
}
\DeclareAcronym{FRW}{
  short = FRW ,
  long = Friedmann-Robertson-Walker ,
  short-plural =  ,
}
\DeclareAcronym{IMR}{
  short = IMR ,
  long = inspiral-merger-ringdown ,
  short-plural =  ,
}
\DeclareAcronym{LISA}{
	short = LISA ,
	long  = Laser Interferometer Space Antenna,
  short-plural =  ,
}
\DeclareAcronym{ET}{
	short = ET ,
	long  = Einstein Telescope,
  short-plural =  ,
}
\DeclareAcronym{CE}{
	short = CE ,
	long  = Cosmic Explorer,
  short-plural =  ,
}
\DeclareAcronym{BBO}{
	short = BBO ,
	long  = Big Bang Observer,
  short-plural =  ,
}
\DeclareAcronym{DECIGO}{
	short = DECIGO ,
	long  = Deci-hertz Interferometer Gravitational wave Observatory,
  short-plural =  ,
}
\DeclareAcronym{ABH}{
	short = ABH ,
	long  = astrophysical black hole,
  short-plural = s ,
}

\DeclareAcronym{PNG}{
	short = PNG ,
	long  = primordial non-Gaussianity ,
  short-plural =  ,
}

\DeclareAcronym{CMB}{
	short = CMB ,
	long  = cosmic microwave background ,
  short-plural =  ,
}
\DeclareAcronym{LSS}{
	short = LSS ,
	long  = large-scale structure ,
  short-plural =  ,
}

\DeclareAcronym{PGW}{
	short = PGW ,
	long  = primordial gravitational wave ,
  short-plural = s ,
}
\DeclareAcronym{SIGW}{
	short = SIGW ,
	long  = scalar-induced gravitational wave ,
  short-plural = s ,
}

\DeclareAcronym{RD}{
	short = RD,
	long  = radiation-dominated ,
  short-plural =  ,
}
\DeclareAcronym{MD}{
	short = MD,
	long  = matter-dominated ,
  short-plural =  ,
}
\DeclareAcronym{eMD}{
	short = eMD,
	long  = early-matter-dominated ,
  short-plural =  ,
}
\DeclareAcronym{SW}{
	short = SW,
	long  = Sachs-Wolfe ,
  short-plural =  ,
}
\DeclareAcronym{ISW}{
	short = ISW,
	long  = integrated Sachs-Wolfe ,
  short-plural =  ,
}

\DeclareAcronym{DM}{
	short = DM,
	long  = dark matter ,
  short-plural =  ,
}

\DeclareAcronym{NANOGrav}{
	short = NANOGrav ,
	long  = North American Nanohertz Observatory for Gravitational Waves ,
  short-plural =  ,
}

\DeclareAcronym{PDF}{
	short = PDF ,
	long  = probability distribution function ,
  short-plural = s ,
}

\DeclareAcronym{SMBH}{
  short = SMBH ,
  long  = supper-massive black hole ,
  short-plural = s ,
}
\DeclareAcronym{SKA}{
	short = SKA ,
	long  = Square Kilometre Array,
  short-plural =  ,
}
\DeclareAcronym{NG15}{
  short = NG15 ,
  long  = NANOGrav 15-year ,
  short-plural =  ,
}

\title{\boldmath Angular bispectrum and trispectrum of scalar-induced gravitational waves:\\ all contributions from primordial non-Gaussianity $f_\mathrm{NL}$ and $g_\mathrm{NL}$}

\author[a,b]{Jun-Peng Li,}
\emailAdd{lijunpeng@ihep.ac.cn}
\author[a]{Sai Wang\footnote{Corresponding author},} 
\emailAdd{wangsai@ihep.ac.cn}
\author[c]{Zhi-Chao Zhao,}
\emailAdd{zhaozc@cau.edu.cn}
\author[d,e,f]{Kazunori Kohri}
\emailAdd{kohri@post.kek.jp}

\affiliation[a]{Theoretical Physics Division, Institute of High Energy Physics, Chinese Academy of Sciences, 19B Yuquan Road, Shijingshan District, Beijing 100049, China}
\affiliation[b]{School of Physics, University of Chinese Academy of Sciences, 19A Yuquan Road, Shijingshan District, Beijing 100049, China}
\affiliation[c]{Department of Applied Physics, College of Science, China Agricultural University, 17 Qinghua East Road, Haidian District, Beijing 100083, China}
\affiliation[d]{Division of Science, National Astronomical Observatory of Japan (NAOJ), and SOKENDAI, 2-21-1 Osawa, Mitaka, Tokyo 181-8588, Japan}
\affiliation[e]{Theory Center, IPNS, and QUP (WPI), KEK, 1-1 Oho, Tsukuba, Ibaraki 305-0801, Japan}
\affiliation[f]{Kavli IPMU (WPI), UTIAS, The University of Tokyo, Kashiwa, Chiba 277-8583, Japan}

\abstract{
Studying the primordial non-Gaussianity of inflationary perturbations is crucial for testing the inflation paradigm of the early universe. In this work, we conduct a comprehensive analysis of the angular bispectrum and trispectrum of scalar-induced gravitational waves (SIGWs) in the presence of local-type primordial non-Gaussianity parameterized by $f_\mathrm{NL}$ and $g_\mathrm{NL}$, deriving their semi-analytical formulae for the first time. Our findings indicate that it is the presence of primordial non-Gaussianity that leads to a non-Gaussian SIGW background, suggesting that the angular bispectrum and trispectrum of SIGWs could serve as probes of the primordial non-Gaussianity. Our numerical results further illustrate that $f_\mathrm{NL}$ and $g_\mathrm{NL}$ exert significant impacts on the spectral amplitudes, potentially reaching up to $10^{-5}$ for the former and $10^{-8}$ for the latter. In particular, we demonstrate that the angular bispectrum and trispectrum exhibit characteristic dependence on the angular multipoles and frequency bands. They hold potentials to be measured by gravitational-wave detectors that may advance our understanding of the origin of the universe. 
}

\pgfplotsset{compat=1.18}
\begin{document}
 
\maketitle
\flushbottom

\section{Introduction}\label{sec:intro}

Primordial non-Gaussianity stands for the deviation from Gaussian distribution of the primordial curvature perturbations, thus characterizing the dynamics of the early universe during inflation \cite{Maldacena:2002vr,Bartolo:2004if,Allen:1987vq,Bartolo:2001cw,Acquaviva:2002ud,Bernardeau:2002jy,Chen:2006nt}. 
Though the standard inflation paradigm predicts the nearly Gaussian perturbations \cite{Salopek:1988qh,Lyth:1998xn}, many other inflation models hold no brief for this prediction \cite{Meerburg:2019qqi}. 
In particular, the primordial bispectrum is characterized by an amplitude parameter $\fnl$, while the primordial trispectrum by $\gnl$ and so on \cite{Smidt:2010ra}.
Constraints on the primordial non-Gaussianity have been obtained through observations of \ac{CMB} \cite{Planck:2019kim} and \ac{LSS} \cite{Castorina:2019wmr,Biagetti:2020skr}. 
However, these observations are only responsive to the cosmological perturbations on large scales that are comparable to the overall observational patch of the universe, leaving the perturbations on smaller scales to be insensitive. 
Since modes of smaller wavelength reentered the horizon earlier, the universe before the last-scattering surface would have been opaque to the electromagnetic probes \cite{Dodelson:2003ft}. 
Therefore, to conquer the above challenge, it is necessary to develop alternative probes that can effectively convey information from the early universe to our detectors. 
Given that Einstein's general relativity predicts that \acp{GW} propagate without dissipation \cite{Bartolo:2018igk,Flauger:2019cam}, we expect them to be a valuable new messenger of the early universe on small scales, in contrast to the traditional observations such as \ac{CMB} and \ac{LSS} on large scales.

The \acp{SIGW} are sensitive to the local-type primordial non-Gaussianity on small scales, making them potential indicators of this non-Gaussianity. 
Theoretically, they were produced through non-linear processes by the linear inflationary perturbations that reentered the Hubble horizon in the early universe \cite{Ananda:2006af,Baumann:2007zm,Espinosa:2018eve,Kohri:2018awv,Mollerach:2003nq,Assadullahi:2009jc,Domenech:2021ztg}. 
Taking into account $\fnl$ and $\gnl$ can lead to significant alterations by several orders of magnitude in both the energy-density fraction spectrum and the angular power spectrum \cite{Li:2023xtl}, especially within the projected sensitivity regimes of future or futuristic gravitational-wave detection programs \cite{Baker:2019nia,Smith:2019wny,Hu:2017mde,Wang:2021njt,Ren:2023yec,TianQin:2015yph,TianQin:2020hid,Seto:2001qf,Kawamura:2020pcg,Crowder:2005nr,Smith:2016jqs,Capurri:2022lze,Hobbs:2009yy,Demorest:2012bv,Kramer:2013kea,Manchester:2012za,Sesana:2008mz,Thrane:2013oya,Janssen:2014dka,2009IEEEP..97.1482D,Weltman:2018zrl,Moore:2014lga,KAGRA:2013rdx,LIGOScientific:2014pky,Chen:2021nxo}. 
This suggests the potential for measuring primordial non-Gaussianity through the search for an anisotropic \ac{SIGW} background.
Other related works can be found in Refs.~\cite{Nakama:2016gzw,Garcia-Bellido:2017aan,Adshead:2021hnm,Ragavendra:2021qdu,Abe:2022xur,Yuan:2023ofl,Yu:2023jrs,Ragavendra:2020sop,Garcia-Saenz:2022tzu,Cai:2018dig,Unal:2018yaa,Atal:2021jyo,Yuan:2020iwf,Zhang:2021rqs,Yuan:2021qgz,Lin:2021vwc,Chen:2022dqr,Cai:2019amo,Cai:2019elf} and Refs.~\cite{Bartolo:2019zvb,Li:2023qua,Wang:2023ost}, respectively. 
Additionally, in conjunction with the production of scalar-induced gravitational waves, these scalar perturbations may gravitationally collapse into \acp{PBH} \cite{Hawking:1971ei,Sasaki:2018dmp}, which are believed to be a reasonable candidate for cold dark matter (e.g., see review in Ref. \cite{Carr:2020xqk,Carr:2023tpt}).
The characteristic mass function of \acp{PBH} can also be significantly altered due to presence of the primordial non-Gaussianity \cite{Bullock:1996at,Byrnes:2012yx,Young:2013oia,Franciolini:2018vbk,Passaglia:2018ixg,Atal:2018neu,Atal:2019cdz,Taoso:2021uvl,Meng:2022ixx,Chen:2023lou,Kawaguchi:2023mgk,Fu:2020lob,Inomata:2020xad,Young:2014ana,Choudhury:2023kdb,Garcia-Bellido:2017aan,Nakama:2016gzw,Ferrante:2022mui,Green:2020jor,Carr:2020gox,Escriva:2022duf,Escriva:2022pnz,Ezquiaga:2019ftu,Kehagias:2019eil,Cai:2021zsp,Cai:2022erk,Yi:2020cut,Zhang:2021vak}.

Higher-order statistics of the energy-density perturbations, the angular bispectrum and trispectrum in particular, characterize the non-Gaussianity of \ac{SIGW} background. 
While the angular power spectrum is a Fourier counterpart of the two-point correlator, the angular bispectrum and trispectrum correspond to the three- and four-point correlators, which are a natural extrapolation of skewness and kurtosis for a random variable.  
They would be highly valuable in extracting additional physical information about inflation, serving as another means to probe the primordial non-Gaussianity. 
It is reasonably anticipated that \ac{GW} background produced by the Gaussian inflationary perturbations is also Gaussian. 
Conversely, the non-Gaussianity of \ac{SIGW} background can result from non-Gaussian inflationary perturbations, indicating that non-vanishing angular bispectrum and trispectrum would be a smoking gun of the primordial non-Gaussianity. 
Recently, the angular bispectrum of a generic cosmological \ac{GW} background has been investigated in Refs.~\cite{Bartolo:2019oiq,Bartolo:2019yeu}, while the study about angular trispectrum of \ac{GW} background remains a blank. 
Additional studies on the non-Gaussianity of \acp{SIGW} are available in Refs. \cite{Bartolo:2018qqn,Bartolo:2018evs,Bartolo:2018rku,Jiang:2024dxj,Zhu:2024xka}. 
Moreover, the angular bispectrum of \acp{SIGW} has been evaluated for the first time in Ref.~\cite{Bartolo:2019zvb}, though only $\fnl$ was crudely taken into account.

In this work, we will perform a comprehensive analysis of the angular bispectrum and trispectrum of \acp{SIGW} by simultaneously taking into account the local-type primordial non-Gaussianity parameterized by $\fnl$ and $\gnl$. 
Following a diagrammatic approach \cite{Li:2023xtl}, we will study the non-Gaussianity of \ac{SIGW} background via including higher-order energy-density perturbations in \acp{SIGW}. 
We will derive semi-analytic formulae for the angular bispectrum and trispectrum and further find the consistency relations between them with the angular power spectrum, respectively. 
In a numerical manner, we will demonstrate significant influence of the primordial non-Gaussianity on both their spectral amplitudes and profiles. 
Moreover, we will prove the non-vanishing angular bispectrum or 
trispectrum of \acp{SIGW} to be a smoking gun of the primordial non-Gaussianity. 
We wish this work to lay a solid foundation for observational studies of the early-universe physics with the cutting-edge \ac{GW} probe, complementing other traditional probes such as \ac{CMB}, \ac{LSS}, and so on.

The paper is organized as follows. 
We briefly review the energy-density fraction spectrum of \acp{SIGW} in Section~\ref{sec:ogw}. 
We introduce the angular correlation functions of the density contrasts of \acp{SIGW} in Section~\ref{sec:dgw}. 
We derive the semi-analytic formulae for the angular bispectrum and trispectrum of \acp{SIGW} in Section~\ref{sec:bis}, and demonstrate the corresponding numerical results in Section~\ref{sec:num}. 
Conclusions and discussion are shown in Section~\ref{sec:conclusion}.

\section{Effects of primordial non-Gaussianity on the SIGW energy density}\label{sec:ogw}


The \ac{SIGW} background arises from the non-linear interactions of primordial curvature perturbations, with the primordial non-Gaussianity anticipated to leave significant imprints on it. 
Apart from the alteration of the averaged energy density spectrum, the specific shape of primordial non-Gaussianity that stands for the coupling between long- and short- wavelength curvature perturbations modulates the distribution of \acp{SIGW}. 
This modulation creates inhomogeneities on large scales, resulting in an anisotropic \ac{SIGW} background in observations.
Furthermore, this modulation also inevitably causes a departure from Gaussian statistic in the \ac{SIGW} background, which can be described by the angular bispectrum and trispectrum, etc. 

First of all, we will review the impacts of the statistics of primordial curvature perturbations, denoted as $\zeta$, on the energy-density fraction spectrum of \acp{SIGW}. 
Schematically, the strain of \acp{SIGW} can be represented as $h_{ij}\sim\zeta^2$ \cite{Ananda:2006af,Baumann:2007zm,Espinosa:2018eve,Kohri:2018awv}, and the corresponding energy density on subhorizon scales is expressed in terms of a four-point correlator of $\zeta$, i.e., $\rho_\uGW \sim {h_{ij,l} h_{ij,l}}\sim\zeta^4$. 
Therefore, both the bispectrum and trispectrum of $\zeta$ can contribute to it. 
In this work, we focus on the local-type primordial non-Gaussianity characterized by $\fnl$ and $\gnl$. 
At a spatial location $\bx$, we express $\zeta$ in terms of their Gaussian components $\zeta_{g}$, namely, 
\cite{Komatsu:2001rj,Okamoto:2002ik,Smidt:2010ra} 
\begin{equation}\label{eq:fnl-gnl-def}
    \zeta (\bx) = \zeta_g (\bx) + \frac{3}{5}\fnl \left[ \zeta_g^2(\bx) - \langle \zeta_g^{2}(\bx) \rangle \right] + \frac{9}{25}\gnl \zeta_g^3(\bx) \ ,
\end{equation}
where the angle brackets stand for an ensemble average. 
Though $\fnl$ and $\gnl$ are scale-independent, our work can be readily extended to a scale-dependent case, which is left to our future works. 
Using Eq.~\ref{eq:fnl-gnl-def}, the multi-point correlators of $\zeta$ can be contracted as the sum of two-point correlator of $\zeta_g$. 
For example, the leading order of the primordial curvature bispectrum can be expressed as $\cO(\fnl)\langle\zeta_g^2\rangle^2$, and the trispectrum as $\cO(\fnl^2)\langle\zeta_g^2\rangle^3 + \cO(\gnl)\langle\zeta_g^2\rangle^3$.
We further decompose $\zeta_{g}$ into short-wavelength modes $\zeta_{gS}$ and long-wavelength modes $\zeta_{gL}$, i.e.,
\cite{Tada:2015noa} 
\begin{equation}\label{eq:S-L-dec}
    \zeta_g=\zeta_{gS}+\zeta_{gL} \ .
\end{equation}
Considering Eqs.~(\ref{eq:fnl-gnl-def},\ref{eq:S-L-dec}), we can express the four-point correlator of $\zeta$ in terms of $\fnl$, $\gnl$, and the dimensionless power spectra of $\zeta_{gX}$ defined as 
\begin{equation}\label{eq:PgS-def}
    \langle \zeta_{gX} (\bq) \zeta_{gX} (\bq') \rangle 
    = \delta^{(3)} (\bq+\bq') \frac{2\pi^2}{q^3} \Delta^2_{X} (q)\ ,
\end{equation}
where $\bq$ denotes the wavevector, and a subscript $_X$ stands for $_S$ and $_L$, respectively. 
For simplicity, we adopt a scale-invariant spectrum for $\Delta_{L}^{2}$, i.e., \cite{Planck:2018vyg} 
\begin{equation}
\Delta_{L}^{2} = A_L \simeq 2.1\times10^{-9}\ .
\end{equation}
We further assume $\Delta^2_{S}$ to be a normal function with respect to $\ln q$, expressed as 
\begin{equation}\label{eq:Lognormal}
    \Delta^2_{S} (q) = \frac{A_S}{\sqrt{2\pi\sigma^2}}\exp\left[-\frac{\ln^2 (q/q_\ast)}{2 \sigma^2}\right]\ ,
\end{equation}
where $\sigma$ represents the spectral width, $q$ is the wavenumber, and $A_S$ denotes the spectral amplitude at the peak wavenumber $q_\ast$. 
We consider a range of $A_S$ spanning from $10^{-4}$ to $10^{-1}$, which is particularly relevant in the context of \ac{PBH} formation scenarios (see reviews in Refs.~\cite{Green:2020jor,Carr:2020gox,Escriva:2022duf} and references therein). 
In the following, we set $\sigma = 1$ for the sake of simplicity, but other values can be readily accommodated if necessary. 
It is worth noting that the perturbativity imposes constraints on the model parameters, specifically 
{  $1 > 3|\fnl| \sqrt{A_S}/5 + 9|\gnl| A_S/25$ and $3|\fnl| \sqrt{A_S}/5 > 9|\gnl| A_S/25$}.

The \acp{SIGW} can be divided into a homogeneous and isotropic background and the fluctuations on it, which are inhomogeneous and anisotropic. 
The former is described by the energy-density fraction spectrum $\bar{\Omega}_\uGW$, while the latter is described by the density contrast $\delta_\uGW$, for which the statistical information is given by their correlation functions, such as the angular power spectrum, bispectrum, trispectrum, and so on. 
The contribution from $\zeta_{gL}$ to $\bar{\Omega}_\uGW$ is negligible due to $A_L \ll A_S$. 
In contrast, $\delta_\uGW$ is relevant with $\zeta_{gL}$ (or $A_L$ equivalently) as the fluctuations arise from the couplings between $\zeta_{gL}$ and $\zeta_{gS}$.
We briefly review $\bar{\Omega}_\uGW$ in the following paragraph and leave the study of $\delta_\uGW$ to the subsequent sections.

The energy-density fraction spectrum $\bar{\Omega}_{\uGW}$ of \acp{SIGW} is defined by $\bar{\rho}_\uGW (\eta) = \rho_\uc(\eta) \int \bar{\Omega}_{\uGW} (\eta,q) \ud\ln q$ \cite{Maggiore:1999vm}, where $\rho_\uc = 3 \cH^2 / (8 \pi G)$ represents the critical energy density of the universe at the conformal time $\eta$, $\cH$ denotes the conformal Hubble parameter, and the overbar signifies physical quantities at the background level. 
As a result, it is also associated with the four-point correlator of $\zeta$, namely $\bar{\Omega}_{\uGW} \sim \langle\zeta^4\rangle$. 
Based on Eqs.~(\ref{eq:fnl-gnl-def},\ref{eq:S-L-dec},\ref{eq:Lognormal}), we decompose it into nine components of the form \cite{Li:2023xtl} 
\begin{equation}
\bar{\Omega}_\uGW^{(a,b)} (\eta_\uin,q) \propto \left(\frac{3}{5}\fnl\right)^{2a} \left(\frac{9}{25}\gnl\right)^b  A_{S}^{a+b+2}  \ , 
\end{equation}
where $a$ and $b$ are natural numbers satisfying the constraint $2 a + b \leq 4$. 
They have been explicitly calculated via following the diagrammatic approach, which has been extensively employed in the literature \cite{Adshead:2021hnm,Ragavendra:2021qdu,Abe:2022xur,Li:2023qua}. 
To be specific, they correspond to the diagrams illustrated in Table~2 and Figure~3 of Ref.~\cite{Li:2023xtl}, with the numerical results displayed in Figure~5 of Ref.~\cite{Li:2023xtl}. 
In particular, $\bar{\Omega}_\uGW^{(0,0)}$ denotes the energy-density fraction spectrum of \acp{GW} produced by the Gaussian scalar perturbations, as demonstrated by the semi-analytic calculation in Refs.~\cite{Espinosa:2018eve,Kohri:2018awv}. 
In summary, we get 
\begin{eqnarray}\label{eq:Omegabar-total}
    \bar{\Omega}_\uGW = \bar{\Omega}_\uGW^{(0,0)} + \bar{\Omega}_\uGW^{(0,1)} + \bar{\Omega}_\uGW^{(1,0)} + \bar{\Omega}_\uGW^{(0,2)} + \bar{\Omega}_\uGW^{(1,1)} + \bar{\Omega}_\uGW^{(2,0)} + \bar{\Omega}_\uGW^{(0,3)} + \bar{\Omega}_\uGW^{(1,2)} + \bar{\Omega}_\uGW^{(0,4)}\ .
\end{eqnarray}
As mentioned above, the contributions from both $\fnl$ and $\gnl$ to $\bar{\Omega}_{\uGW} (\eta_\uin,q)$ have been comprehensively analyzed in our existing work \cite{Li:2023xtl}, with other studies available in Refs.~\cite{Li:2023qua,Wang:2023ost,Nakama:2016gzw,Garcia-Bellido:2017aan,Adshead:2021hnm,Ragavendra:2021qdu,Abe:2022xur,Yuan:2023ofl,Cai:2018dig,Unal:2018yaa,Atal:2021jyo,Yuan:2020iwf,Zhang:2021rqs,Yuan:2021qgz,Lin:2021vwc,Chen:2022dqr,Ragavendra:2020sop,Garcia-Saenz:2022tzu,Yu:2023jrs}. 

For a given frequency band $\nu=q/(2\pi)$, the present-day energy-density fraction spectrum $\bar{\Omega}_{\uGW,0} (\nu)$ is proportional to the energy-density fraction spectrum $\bar{\Omega}_{\uGW}$ introduced in Eq.~\eqref{eq:Omegabar-total}. 
It is given by \cite{Wang:2019kaf}
\begin{eqnarray}\label{eq:Omega0}
    \bar{\Omega}_{\uGW,0} (\nu) 
     \simeq  \Omega_{\mathrm{r}, 0} \bar{\Omega}_\uGW (\eta_\uin,q) \ ,
\end{eqnarray}
where $\Omega_{\mathrm{r},0}=9.265 \times 10^{-5}$ is the present-day energy-density fraction of radiation in the universe \cite{Planck:2018vyg}.
In the above equation, we have neglected effects of the effective number of relativistic degrees of freedom on the cosmic scale factor. 
They would not change our main results of this work, but lead to corrections of a factor of at most 2, which can be recovered if needed.

\section{Angular correlation functions}\label{sec:dgw}

We introduce the density contrast to quantify deviations from the homogeneous and isotropic background, given by \cite{Bartolo:2019oiq,Bartolo:2019yeu} 
\begin{equation}\label{eq:delta-def}
    \delta_\uGW (\eta,\bx,\bq) = 4\pi \frac{\omega_\uGW (\eta,\bx,\bq)}{\bar{\Omega}_\uGW (\eta,q)} - 1 \ ,
\end{equation}
where the energy-density full spectrum $\omega_{\uGW}$ is defined by $\rho_\uGW (\eta,\bx) = \rho_\uc(\eta) \int \ud \ln q \, \ud^{2} \hat{\bq} \, \omega_\uGW (\eta,\bx,\bq)$.
Furthermore, the energy density of \acp{SIGW} is explicitly expressed as $\rho_\uGW = \overbar{\partial_l h_{ij} \partial_l h_{ij}} / (128 \pi G a^2)$, where $a(\eta)$ is the scale factor of the universe, the long overbar stands for an oscillation average, and the semi-analytic formula of the \ac{SIGW} strain $h_{ij}(\eta,\bq)$ is available in Refs.~\cite{Espinosa:2018eve,Kohri:2018awv}. 
Therefore, we have 
\begin{equation}\label{eq:omega-h}
     \omega_\uGW (\eta,\bx,\bq)  
     = - \frac{q^3}{48 \cH^2} \int \frac{\ud^3 \bk}{(2\pi)^{3}} e^{i\bk\cdot\bx} 
        \left[\left(\bk-\bq\right) \cdot \bq \right] \overbar{h_{ij}(\eta,\bk-\bq)
        h_{ij}(\eta,\bq)}\ ,
\end{equation}
which depends on both $q$ and the direction $\hat{\bq}$. 
Averaging over a vast number of Hubble horizons around $\bx$ at the production time yields $\langle \omega_\uGW(\eta,\bx,\bq) \rangle_{\bx} = \bar{\Omega}_\uGW(\eta,q) / (4 \pi)$, confirming the condition $\langle \delta_\uGW(\eta,\bx,\bq) \rangle_{\bx} = 0$.

Based on the Boltzmann equation \cite{Contaldi:2016koz,Bartolo:2019oiq,Bartolo:2019yeu}, the present-day density contrast $\delta_{\uGW,0}(\bq)=\delta_{\uGW}(\eta_{0},\bx_{0},\bq)$ incorporates the initial inhomogeneity $\delta_\uGW (\eta_\uin,\bx,\bq)$ and the propagation effects. 
It is given by \cite{Bartolo:2019zvb,Li:2023qua,Li:2023xtl,Wang:2023ost}
\begin{equation}\label{eq:delta-0}
    \delta_{\uGW,0}(\bq) = \delta_\uGW (\eta_\uin,\bx,\bq) + \left[4-n_{\uGW} (\nu)\right] \Phi (\eta_\uin, \bx)\ ,
\end{equation}
where the metric scalar perturbation $\Phi(\eta_\uin,\bx)$ represents the \ac{SW} effect \cite{Sachs:1967er}, and the energy-density fraction spectral index, defined as 
\begin{equation}\label{eq:ngw-def}
    n_{\uGW} (\nu) = \frac{\partial\ln \bar{\Omega}_{\uGW,0} (\nu)}{\partial\ln \nu} \ ,
\end{equation}
is approximated to be time-independent owing to Eq.~\eqref{eq:Omega0}. 
Here, we disregard the \ac{ISW} effect {  and the gravitational lensing effect. 
The \ac{ISW} effect is relatively less significant than the \ac{SW} effect \cite{Bartolo:2019zvb}, while the gravitational lensing effect is of higher order compared with the \ac{SW} effect \cite{Bartolo:2019oiq,Bartolo:2019yeu} and is expected to primarily impact $\delta_{\uGW,0}$ at higher multipoles \cite{Dodelson:2003ft} }. 
Nevertheless, it is straightforward to extend our analysis to include all of them.

The angular correlators encode statistical information of the inhomogeneity and anisotropy of \acp{SIGW}. 
We introduce $\delta_{\uGW,0,\ell m} (2\pi\nu)$ to represent the decomposition of $\delta_{\uGW,0} (\bq)$ in terms of spherical harmonics, as given by 
\begin{equation}\label{eq:spher-harm}
    \delta_{\uGW,0}(\bq) = \sum_{\ell m} \delta_{\uGW,0,\ell m}(q) Y_{\ell m}(\hat{\bq})\ . 
\end{equation}
The two-point correlator gives rise to the reduced angular power spectrum, i.e., 
\begin{equation}\label{eq:Ctilde-def}
    \left\langle\delta_{\uGW,0,\ell m}(2\pi\nu) \delta_{\uGW,0,\ell' m'}^\ast(2\pi\nu)\right\rangle
    = \delta_{\ell \ell'} \delta_{mm'} \tilde{C}_\ell (\nu)\ ,
\end{equation}
which has been extensively investigated in Refs. \cite{Bartolo:2019zvb,Li:2023qua,Li:2023xtl,ValbusaDallArmi:2020ifo,Dimastrogiovanni:2021mfs,LISACosmologyWorkingGroup:2022kbp,LISACosmologyWorkingGroup:2022jok,Unal:2020mts,Malhotra:2020ket,Carr:2020gox,Cui:2023dlo,Malhotra:2022ply,ValbusaDallArmi:2023nqn,LISACosmologyWorkingGroup:2023njw}. 
In this work, we introduce the (reduced) angular bispectrum and trispectrum to capture the non-Gaussianity of \ac{GW} background, analogue to \ac{CMB}. 

The angular bispectrum for the rotation-invariant \ac{GW} background is factorized through the three-point correlator, i.e., 
\begin{eqnarray}\label{eq:btilde-def}
    \left\langle\prod_{i=1}^3\delta_{\uGW,0,\ell_i m_i}(2\pi\nu)\right\rangle 
    = \Glm 
    \tilde{b}_{\ell_1 \ell_2 \ell_3} (\nu)\ ,
\end{eqnarray} 
where the Gaunt integral $\Glm$ arises from the assumption of statistical isotropy and parity invariance, as expressed in terms of the Wigner 3-$j$ symbols, i.e., 
\begin{equation}
    \Glm 
    = h_{\ell_1 \ell_2 \ell_3} 
    \begin{pmatrix}
         \ell_1 & \ell_2 & \ell_3 \\
         m_1    & m_2    & m_3   
    \end{pmatrix}\ ,
\end{equation}
where we define 
\begin{equation}\label{eq:hlll}
    h_{\ell_1 \ell_2 \ell_3} 
    = \sqrt{\frac{(2\ell_1 + 1)(2\ell_2 + 1)(2\ell_3 + 1)}{4\pi}}
    \begin{pmatrix}
         \ell_1 & \ell_2 & \ell_3 \\
         0      & 0      & 0     
    \end{pmatrix}\ .
\end{equation}
Moreover, we recall that the tetrahedral domain of multipole triplets $\{\ell_1,\ell_2,\ell_3\}$ satisfies both the triangular inequalities ($\ell_1 \leq \ell_2 + \ell_3$, $\ell_2 \leq \ell_1 + \ell_3$, $\ell_3 \leq \ell_1 + \ell_2$, and $m_1 + m_2 + m_3 = 0$) and the parity condition ($\ell_1 + \ell_2 + \ell_3 = 2n$, $n \in \mathbb{N}$), which are identical to those in the study of \ac{CMB} \cite{Komatsu:2001ysk,Planck:2013wtn}.

Going to higher-order statistic, the four-point correlator defines the angular trispectrum. 
Analogous to that of \ac{CMB} \cite{Hu:2001fa,Komatsu:2010hc,Kogo:2006kh}, the angular trispectrum of a rotationally invariant \ac{GW} background takes the following form 
\begin{eqnarray}\label{eq:Tl-def}
    \left\langle\prod_{i=1}^4 \delta_{\uGW,0,\ell_i m_i}(2\pi \nu)\right\rangle
    &=& \sum_{LM} (-1)^M 
    \begin{pmatrix}
         \ell_1 & \ell_2 & L \\
         m_1    & m_2    & -M   
    \end{pmatrix}
    \begin{pmatrix}
         \ell_3 & \ell_4 & L \\
         m_3    & m_4    & M   
    \end{pmatrix}
    T^{\ell_1 \ell_2}_{\ell_3 \ell_4} (L,\nu) \ .
\end{eqnarray}
where $T^{\ell_1 \ell_2}_{\ell_3 \ell_4} (L,\nu)$ stands for the angular averaged trispectrum. 
The Wigner 3-$j$ symbols imply that the quadrilateral $\{\ell_1,\ell_2,\ell_3,\ell_4\}$ is divided into two triangles ($\{\ell_1,\ell_2,L\}$ and $\{\ell_3,\ell_4,L\}$) by a diagonal, which results in the triangular conditions $|\ell_1 - \ell_2| < L < \ell_1 + \ell_2$ and $|\ell_3 - \ell_4| < L < \ell_3 + \ell_4$ and $m_1 + m_2 - M = m_3 + m_4 + M  = 0$. 
Moreover, $\ell_1 + \ell_2 + \ell_3 + \ell_4$ are required to be even as the result of parity invariance. 
Further, $T^{\ell_1 \ell_2}_{\ell_3 \ell_4} (L,\nu)$ can be decomposed into the disconnected part, denoted as ${T_G}^{\ell_1 \ell_2}_{\ell_3 \ell_4} (L,\nu)$, and the connected part, denoted as ${T_c}^{\ell_1 \ell_2}_{\ell_3 \ell_4} (L,\nu)$, i.e., 
\begin{equation}
    T^{\ell_1 \ell_2}_{\ell_3 \ell_4} (L,\nu) 
    = {T_G}^{\ell_1 \ell_2}_{\ell_3 \ell_4} (L,\nu) + {T_c}^{\ell_1 \ell_2}_{\ell_3 \ell_4} (L,\nu)\ .
\end{equation}
The former corresponds to the Gaussian case, and thus can be expressed in terms of $\tilde{C}_\ell (\nu)$, namely,  
\begin{eqnarray}\label{eq:TGl-def}
    {T_G}^{\ell_1 \ell_2}_{\ell_3 \ell_4} (L,\nu) 
    &=& (-1)^{\ell_1 + \ell_3} \sqrt{(2\ell_1 + 1)(2\ell_3 + 1)} \tilde{C}_{\ell_1} (\nu) \tilde{C}_{\ell_3} (\nu) \delta_{\ell_1 \ell_2} \delta_{\ell_3 \ell_4} \delta_{L 0} \nonumber\\ 
    && + (2 L + 1) \tilde{C}_{\ell_1} (\nu) \tilde{C}_{\ell_2} (\nu) \Bigl[(-1)^{\ell_1 + \ell_2 + L} \delta_{\ell_1 \ell_3} \delta_{\ell_2 \ell_4} + \delta_{\ell_1 \ell_4} \delta_{\ell_2 \ell_3}\Bigr]\ .
\end{eqnarray}
In contrast, the later features the non-Gaussianity of \ac{SIGW} background. 
By using the Wigner 6-$j$ symbol to account for the permutation symmetry, we can explicitly express it in the form of 
\begin{eqnarray}\label{eq:Pl-def}
    {T_c}^{\ell_1 \ell_2}_{\ell_3 \ell_4} (L,\nu) 
    &=& P^{\ell_1 \ell_2}_{\ell_3 \ell_4} (L,\nu) 
    + (2L + 1) \sum_{L'} 
    \Biggl[
        (-1)^{\ell_2 + \ell_3}
        \begin{Bmatrix}
            \ell_1 & \ell_2 & L' \\
            \ell_4 & \ell_3 & L
        \end{Bmatrix} 
        P^{\ell_1 \ell_3}_{\ell_2 \ell_4} (L',\nu) \nonumber\\ 
        && \hphantom{P^{\ell_1 \ell_2}_{\ell_3 \ell_4} (L,\nu) 
    + (2L + 1) \sum_{L'}}+ (-1)^{L + L'}
        \begin{Bmatrix}
            \ell_1 & \ell_2 & L' \\
            \ell_3 & \ell_4 & L
        \end{Bmatrix} 
        P^{\ell_1 \ell_4}_{\ell_3 \ell_2} (L',\nu)
    \Biggr]\ ,
\end{eqnarray}
where $P^{\ell_1 \ell_2}_{\ell_3 \ell_4} (L,\nu)$ is written in terms of the reduced angular trispectrum $t^{\ell_1 \ell_2}_{\ell_3 \ell_4} (L,\nu)$, namely, 
\begin{equation}\label{eq:tl-def}
    P^{\ell_1 \ell_2}_{\ell_3 \ell_4} (L,\nu) 
    = t^{\ell_1 \ell_2}_{\ell_3 \ell_4} (L,\nu) 
    + (-1)^{\ell_1 + \ell_2 + L} t^{\ell_2 \ell_1}_{\ell_3 \ell_4} (L,\nu)
    + (-1)^{\ell_3 + \ell_4 + L} t^{\ell_1 \ell_2}_{\ell_4 \ell_3} (L,\nu)
    + (-1)^{\ell_1 + \ell_2 + \ell_3 + \ell_4 + 2L} t^{\ell_2 \ell_1}_{\ell_4 \ell_3} (L,\nu)\ .
\end{equation}
Notably, due to the parity invariance, the reduced angular trispectrum obeys $t^{\ell_1 \ell_2}_{\ell_3 \ell_4} (L,\nu) = t^{\ell_2 \ell_1}_{\ell_4 \ell_3} (L,\nu)$. 
Moreover, the permutation symmetry requires that it is symmetric against exchange of its upper and lower indices, namely, $t^{\ell_1 \ell_2}_{\ell_3 \ell_4} (L,\nu) = t_{\ell_1 \ell_2}^{\ell_3 \ell_4} (L,\nu)$. 
Obviously, the disconnected part ${T_G}^{\ell_1 \ell_2}_{\ell_3 \ell_4} (L,\nu)$ vanishes if all the four multipoles are not equal, so that the angular averaged trispectrum is just the connected part ${T_c}^{\ell_1 \ell_2}_{\ell_3 \ell_4} (L,\nu)$. 
Due to limitations in the angular resolutions of ongoing and planned \ac{GW} detectors \cite{Baker:2019ync,Capurri:2022lze,Gair:2015hra,Romano:2016dpx,LISACosmologyWorkingGroup:2022kbp,LIGOScientific:2016nwa,LIGOScientific:2019gaw,KAGRA:2021mth,NANOGrav:2023tcn}, we consider only the low multipoles $\ell$ in our present work. 
However, it is straightforward to generalize our analysis to higher multipoles $\ell$ if needed.

\section{Semi-analytic formulae for angular bispectrum and trispectrum}\label{sec:bis}

\begin{figure}[htbp]
    \centering
    \includegraphics[width =0.3 \columnwidth]{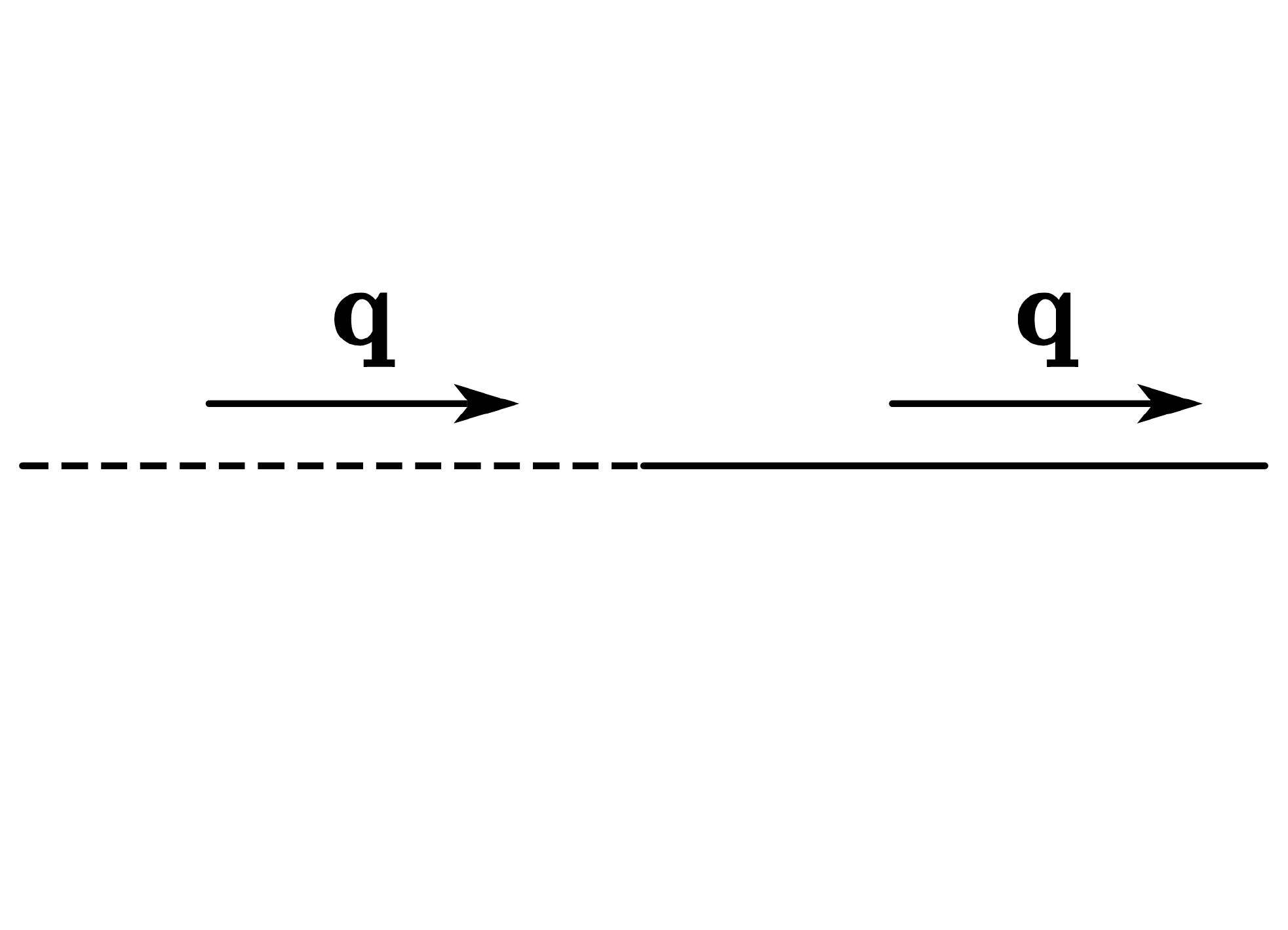}
    \hfil
    \includegraphics[width =0.3 \columnwidth]{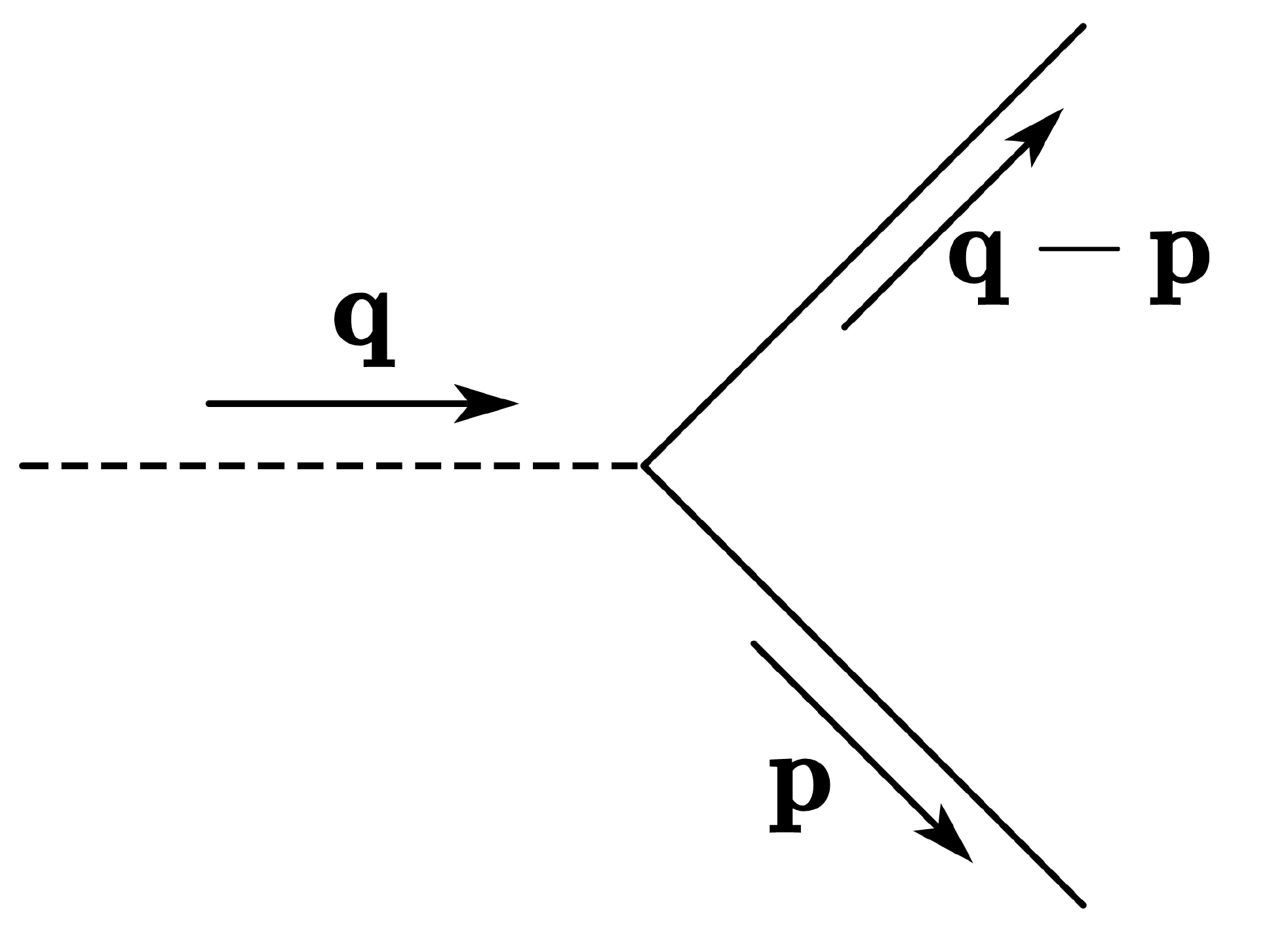}
    \hfil
    \includegraphics[width =0.3 \columnwidth]{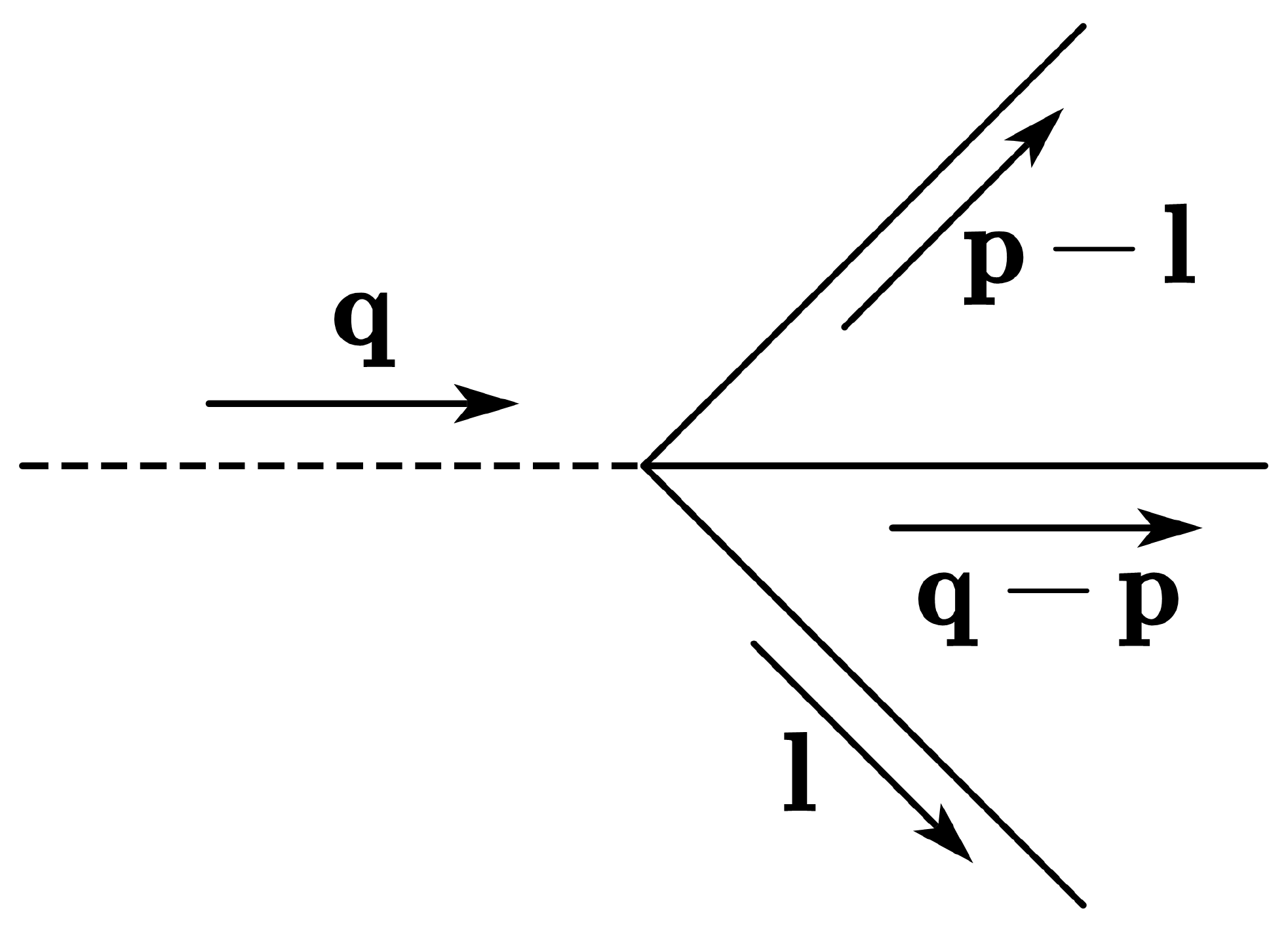}
    \caption{Gaussian-vertex, $\fnl$-vertex, and $\gnl$-vertex.  
    }\label{fig:vertex-type}
\end{figure}

We employ the diagrammatic approach \cite{Li:2023qua,Li:2023xtl} to derive the semi-analytic formulae for the angular bispectrum and trispectrum of \acp{SIGW}. 
As defined in Eq.~\eqref{eq:btilde-def}, $\tilde{b}_{\ell_1 \ell_2 \ell_3}$ and $t^{\ell_1 \ell_2}_{\ell_3 \ell_4} (L,\nu)$ are related with the three- and four-point correlators of $\delta_{\uGW,0}(\bq)$, respectively, which is introduced by Eq.~\eqref{eq:delta-0}. 
While the correlations involving the \ac{SW} effect alone can be straightforwardly evaluated, our calculation of the correlations involving the initial inhomogeneities $\delta_\uGW (\eta_\uin,\bx,\bq)$ would result in multi-point correlations of the form $\langle \omega_\uGW^{3} \rangle \sim \langle h^{6} \rangle \sim \langle \zeta^{12} \rangle$ for $\tilde{b}_{\ell_1 \ell_2 \ell_3}$ and $\langle \omega_\uGW^{4} \rangle \sim \langle h^{8} \rangle \sim \langle \zeta^{16} \rangle$ for $t^{\ell_1 \ell_2}_{\ell_3 \ell_4} (L,\nu)$.
When Eq.~\eqref{eq:fnl-gnl-def} is further considered, it is particularly challenging to evaluate these correlations, since they involve correlations of the order at most $\langle \zeta_{g}^{36} \rangle$ and $\langle \zeta_{g}^{48} \rangle$, respectively. 
As demonstrated by Ref.~\cite{Li:2023xtl}, however, the diagrammatic approach transforms the multi-point correlators of non-Gaussian $\zeta$ into combinations of connections among different $\zeta$ vertices, thus significantly simplifying our calculations. 
Based on Eq.~\eqref{eq:fnl-gnl-def}, the relationship between the primordial non-Gaussian curvature perturbations $\zeta$ up to $\gnl$ order and their Gaussian components $\zeta_g$ can be represented via three types of vertices, as shown in \cref{fig:vertex-type} and named from left to right as the Gaussian-vertex, the $\fnl$-vertex, and the $\gnl$-vertex.

\begin{figure}[htbp]
    \centering
    \includegraphics[width =0.32 \columnwidth]{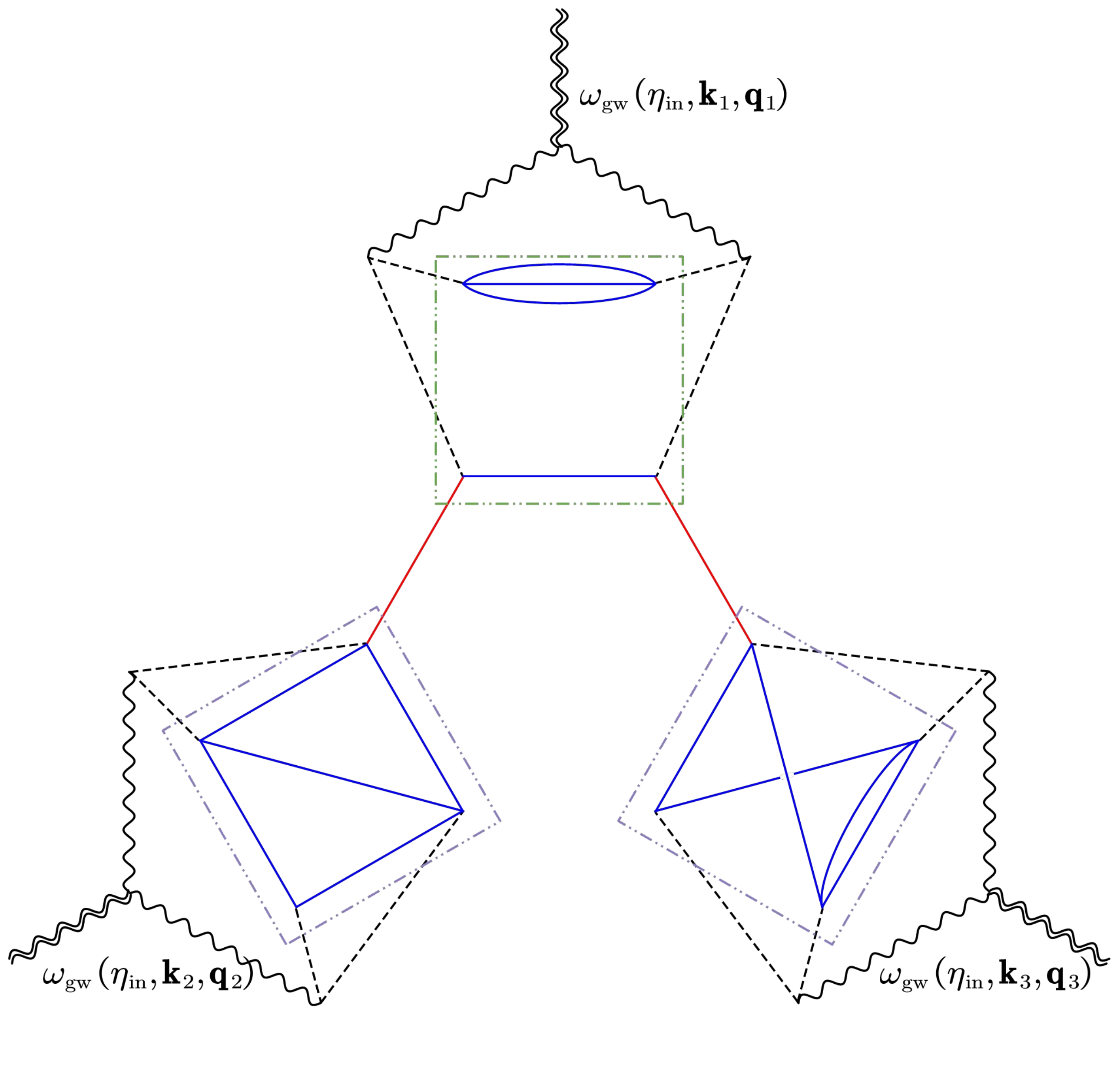}
    \hfil
    \includegraphics[width =0.32 \columnwidth]{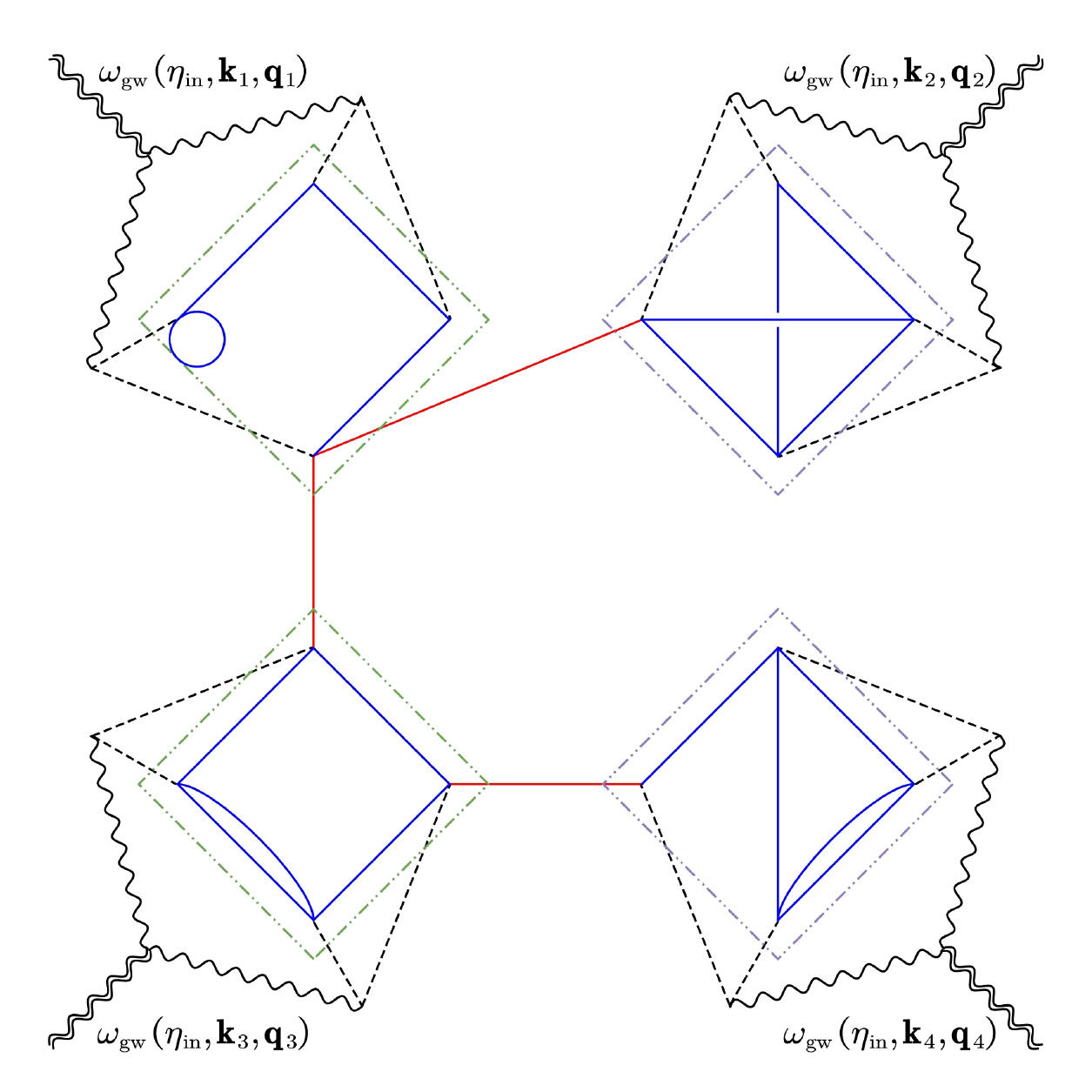}
    \hfil
    \includegraphics[width =0.32 \columnwidth]{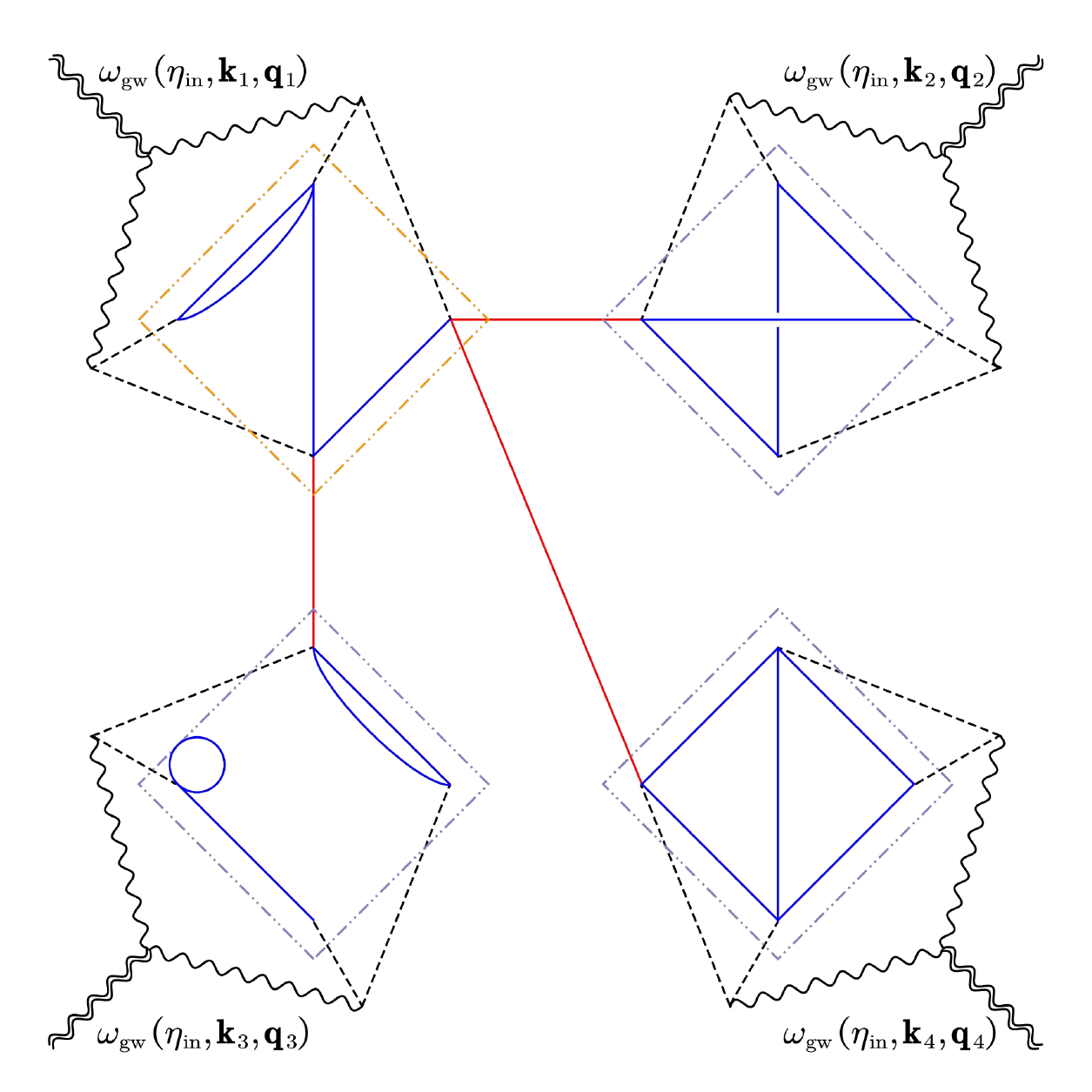}
    \caption{One of the Feynman-like diagrams for the three-point correlator of $\omega_\uGW (\eta_\uin,\bk,\bq)$ (left) and two of the Feynman-like diagrams for the four-point correlator of $\omega_\uGW (\eta_\uin,\bk,\bq)$ (middle and right). 
    Other Feynman-like diagrams can be obtained via replacing the dotted-dashed boxes with certain panels of Figure~3 in Ref.~\cite{Li:2023xtl}, following an explicit approach demonstrated in the context. 
    Here, double wavy lines denote the energy-density full spectrum $\omega_{\uGW}$, single wavy lines denote the \ac{GW} strain $h$, dashed lines denote the scalar transfer functions, and solid lines denote the primordial curvature power spectra for the Gaussian components $\zeta_g$. 
    Especially, the power spectrum of $\zeta_{gS}$ is represented as blue solid lines, while the power spectrum of $\zeta_{gL}$ are represented as red solid lines. 
    }\label{fig:correlator}
\end{figure}

Following the Feynman-like rules in Ref.~\cite{Li:2023xtl}, we are able to depict the connected Feynman-like diagrams for the multi-point correlators of $\omega_\uGW (\eta_\uin,\bk_i,\bq_i)$, which is the Fourier mode of $\omega_\uGW (\eta_\uin,\bx_i,\bq_i)$. 
For the three-point correlator $\langle \prod_{i=1}^3 \omega_\uGW (\eta_\uin,\bk_i,\bq_i) \rangle$ at leading order of $A_L$, one typical Feynman-like diagram is demonstrated in the left panel of \cref{fig:correlator}. 
In the diagram, the initial inhomogeneities $\omega_\uGW$ situate at distinct positions, are connected by ``non-Gaussian bridges'' (denoted as red solid lines), which symbolize the power spectra of $\zeta_{gL}$. 
These connections stand for the redistribution of the \ac{SIGW} energy density on large scales.  
Given that $A_L \ll A_S$, diagrams with more than two ``non-Gaussian bridges'' contribute little to the bispectrum and are discarded. 
In addition, the difference between other Feynman-like diagrams and this one lies in the different ways that the blue or red solid lines connect the vertices within the three dashed boxes. 
Likewise, when considering the four-point correlator $\langle \prod_{i=1}^4 \omega_\uGW (\eta_\uin,\bk_i,\bq_i) \rangle$ at leading order of $A_L$, two typical Feynman-like diagrams on behalf of two types of connections via ``non-Gaussian bridges'' are depicted in the middle and right panels of \cref{fig:correlator}.  
We are able to obtain other Feynman-like diagrams with three  ``non-Gaussian bridges'' through replacing the four dotted-dashed boxes, while the Feynman-like diagrams with more than three ``non-Gaussian bridges'' are discarded as they are negligible.

As demonstrated by our existing work \cite{Li:2023xtl}, it is convenient to expand $\omega_\uGW (\eta_\uin,\bx_i,\bq_i)$ up to $\cO (\zeta_{gL}^3)$ before calculating the correlations, which simplifies the replacement of the four dotted-dashed boxes and avoids computing the cross-correlation of the initial term and the propagation term. 
Using Eqs.~(\ref{eq:fnl-gnl-def},\ref{eq:S-L-dec},\ref{eq:omega-h}) and $h_{ij}\sim\zeta^{2}$, we can express $\omega_\uGW (\eta_\uin,\bx_i,\bq_i)$ schematically through the Wick contractions as follows
\begin{eqnarray}\label{eq:ogwexpand}
    \omega_{\uGW}(\eta_\uin,\bx_i,\bq_i) 
    &\sim& \langle\zeta^4\rangle_{\bx_i} \nonumber\\
    &\sim& \langle\zeta_{S}^4\rangle_{\bx_i} 
    + \mathcal{O}(\zeta_{gL}) \bigl[
        \fnl \langle\zeta_{gS} \zeta_{S}^3\rangle_{\bx_i} +  \gnl \langle\zeta_{gS}^2 \zeta_{S}^3\rangle_{\bx_i}
    \bigr]\nonumber\\ 
    &&+ \mathcal{O}(\zeta_{gL}^2) \bigl[
        \fnl^2 \langle\zeta_{gS}^2 \zeta_{S}^2\rangle_{\bx_i} + \fnl\gnl \langle\zeta_{gS}^3 \zeta_{S}^2\rangle_{\bx_i} + \gnl^2 \langle\zeta_{gS}^4 \zeta_{S}^2\rangle_{\bx_i} + \gnl \langle\zeta_{gS} \zeta_{S}^3\rangle_{\bx_i}
    \bigr]\nonumber\\ 
    &&+ \mathcal{O}(\zeta_{gL}^3) \bigl[
        \fnl^3 \langle\zeta_{gS}^3 \zeta_{S}\rangle_{\bx_i} + \fnl^2\gnl \langle\zeta_{gS}^4 \zeta_{S}\rangle_{\bx_i} + \fnl\gnl^2 \langle\zeta_{gS}^5 \zeta_{S}\rangle_{\bx_i}\nonumber\\
        &&\hphantom{+ \mathcal{O}(\zeta_{gL}^3) \bigl[} 
        + \gnl^3 \langle\zeta_{gS}^6 \zeta_{S}\rangle_{\bx_i} + \fnl\gnl \langle\zeta_{gS}^2 \zeta_{S}^2\rangle_{\bx_i} + \gnl^2 \langle\zeta_{gS}^3 \zeta_{S}^2\rangle_{\bx_i}
    \bigr]\ , 
\end{eqnarray}
where $\zeta_{S}$ represents the short-wavelength modes including both the Gaussian and non-Gaussian components.
Notably, the ensemble average within the observed region centered at $\bx_i$ is necessary, as the observed signal along a line-of-sight is a combination of \acp{SIGW} originating from numerous small-scale regions, owing to limitations in the angular resolution of \ac{GW} detectors.
Furthermore, the leading term $\langle\zeta_{S}^4\rangle_{\bx}$ corresponds to the background $\bar{\Omega}_\uGW / (4 \pi)$ exactly, while others should be combinations of $\bar{\Omega}_\uGW^{(a,b)}$ that contribute to the energy-density fluctuations on large scales, as demonstrated in the following. 

\begin{figure}[htbp]
    \centering
    \includegraphics[width =0.18 \columnwidth]{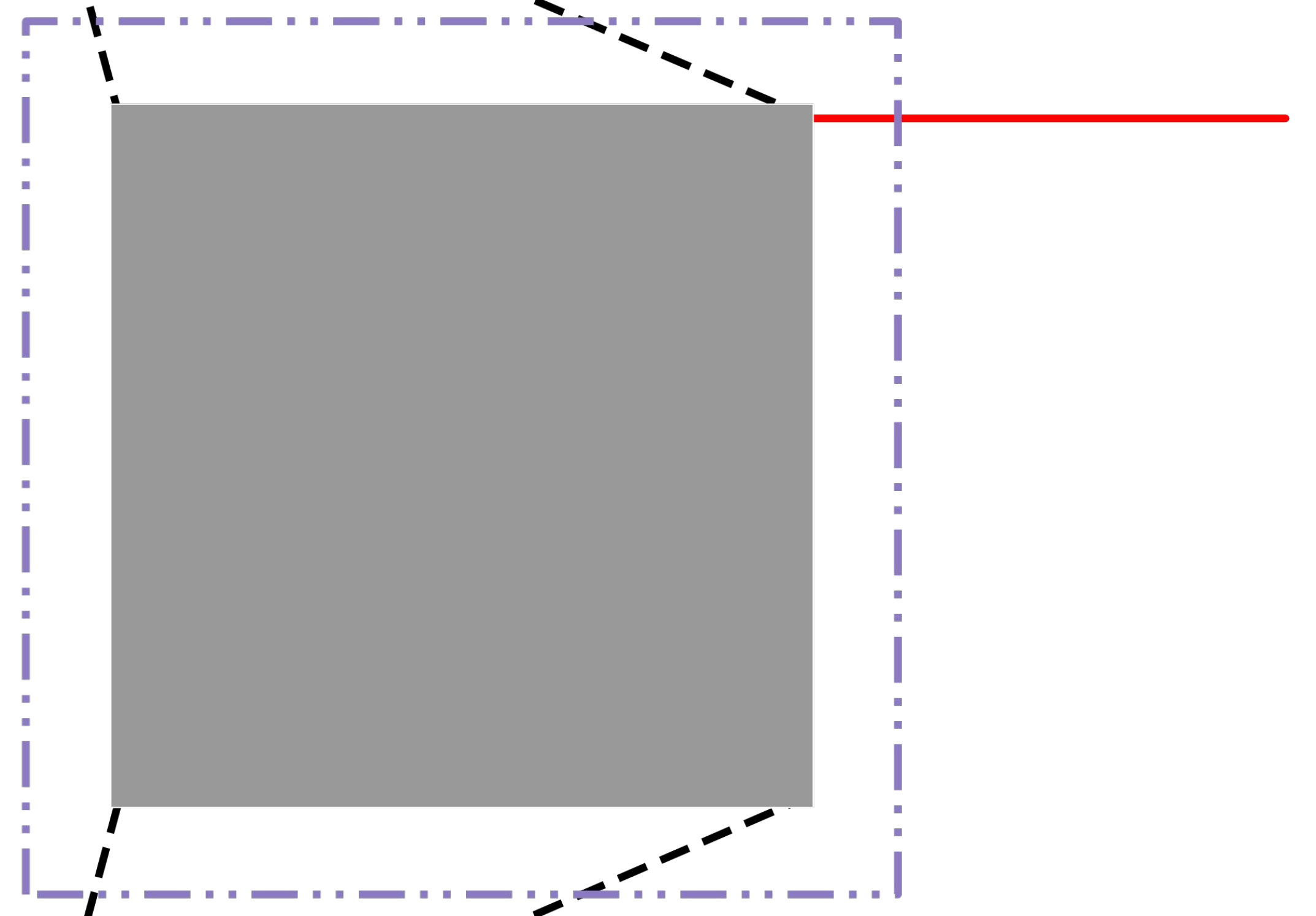}
    \hfil
    \includegraphics[width =0.18 \columnwidth]{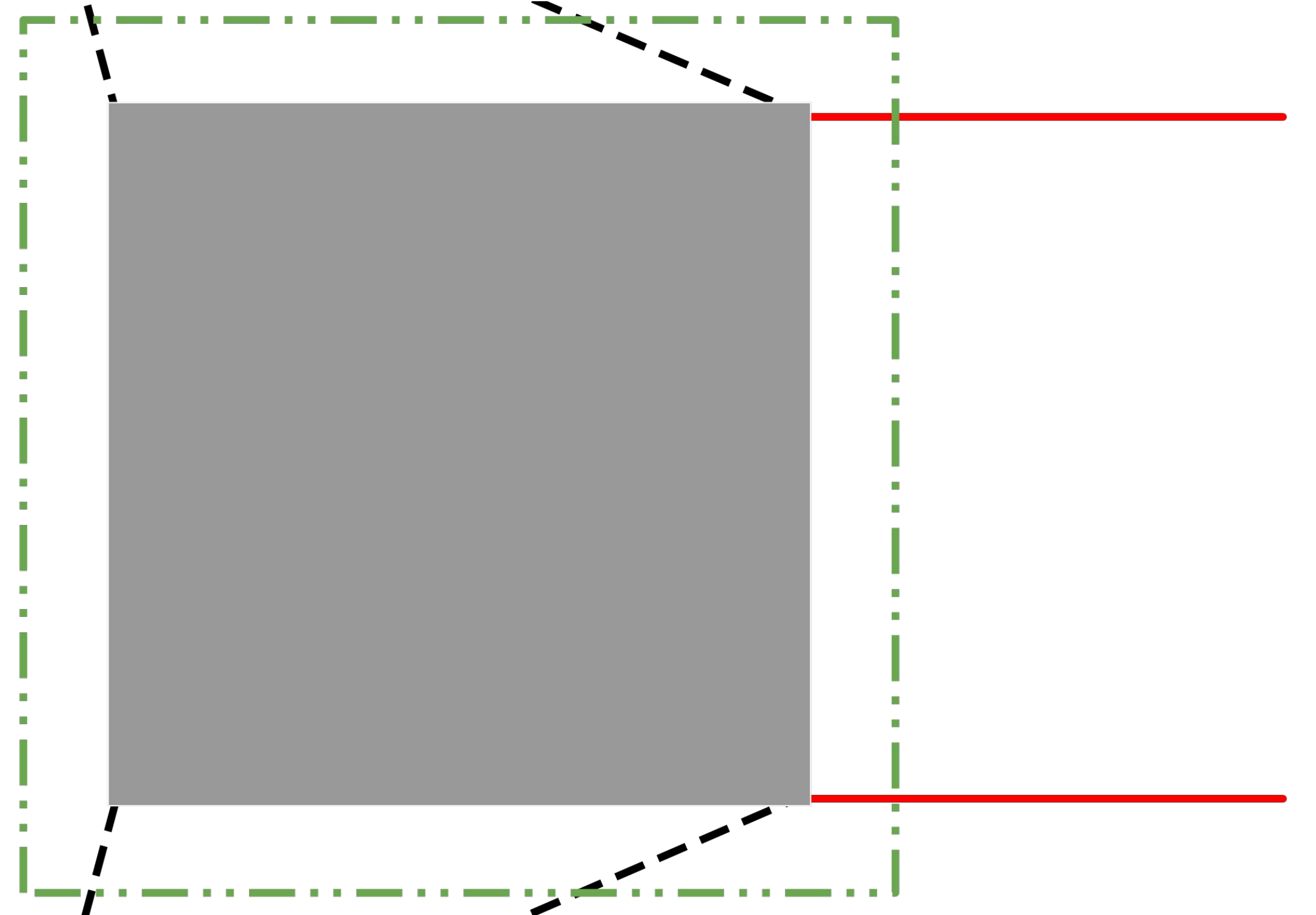}
    \hfil
    \includegraphics[width =0.18 \columnwidth]{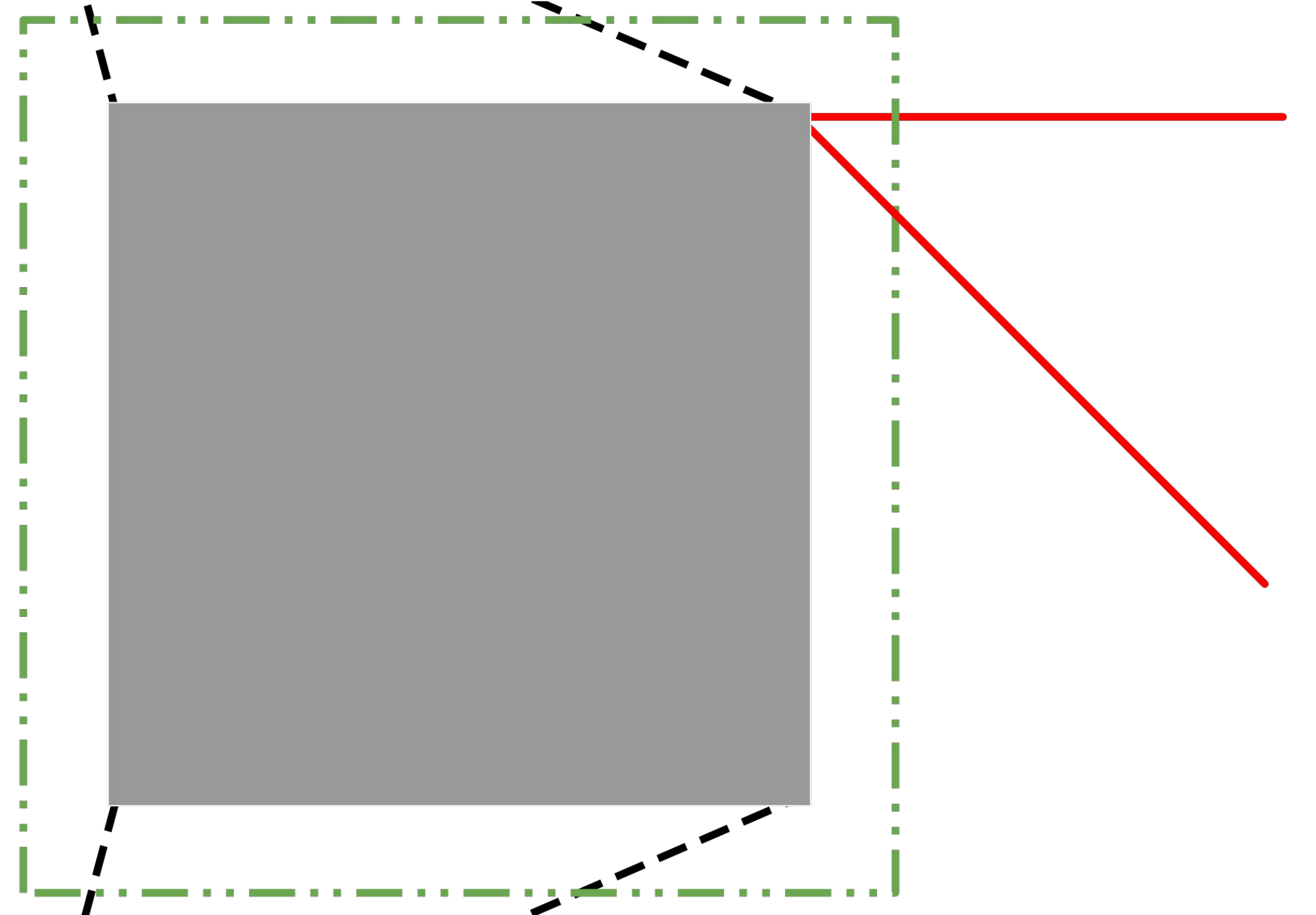}
    \includegraphics[width =0.18 \columnwidth]{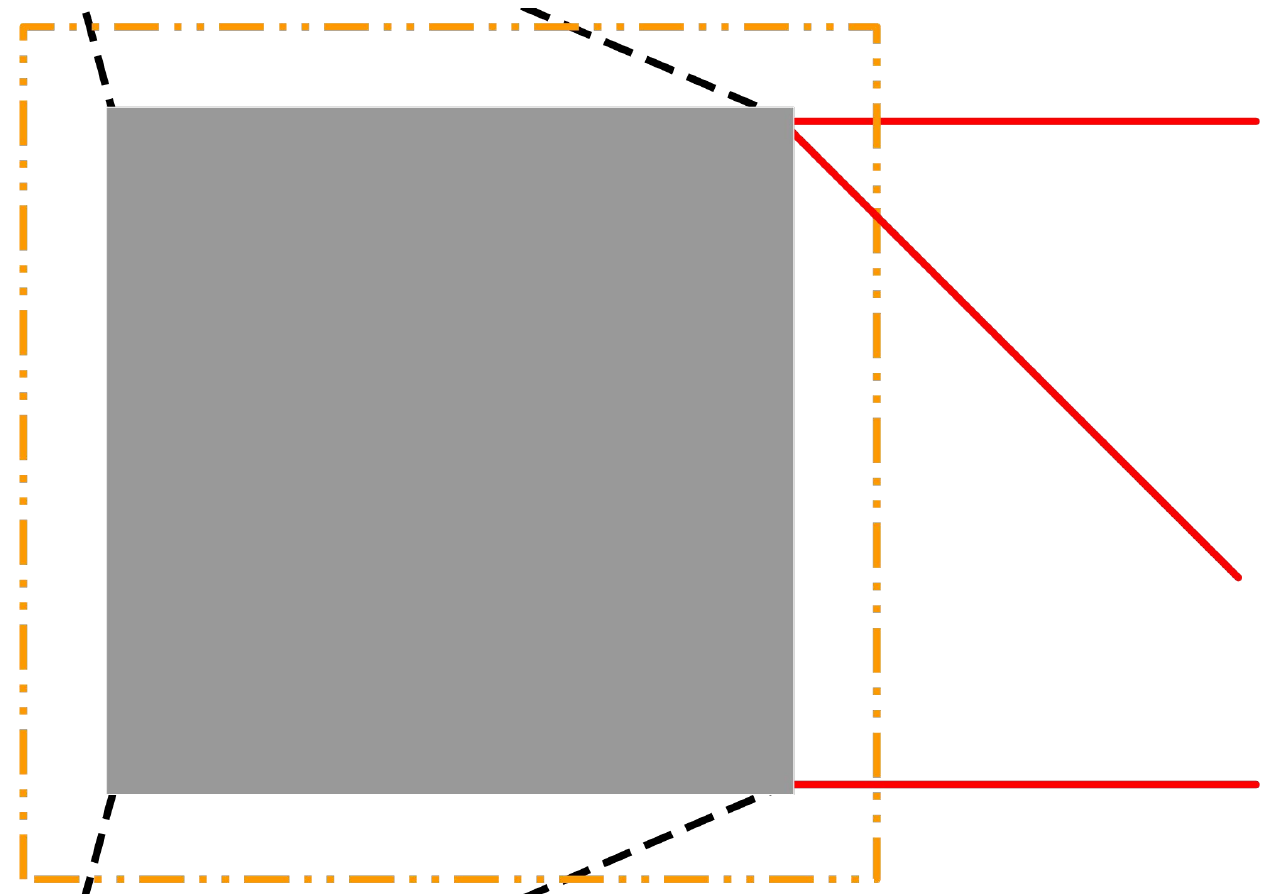}
    \hfil
    \includegraphics[width =0.24 \columnwidth]{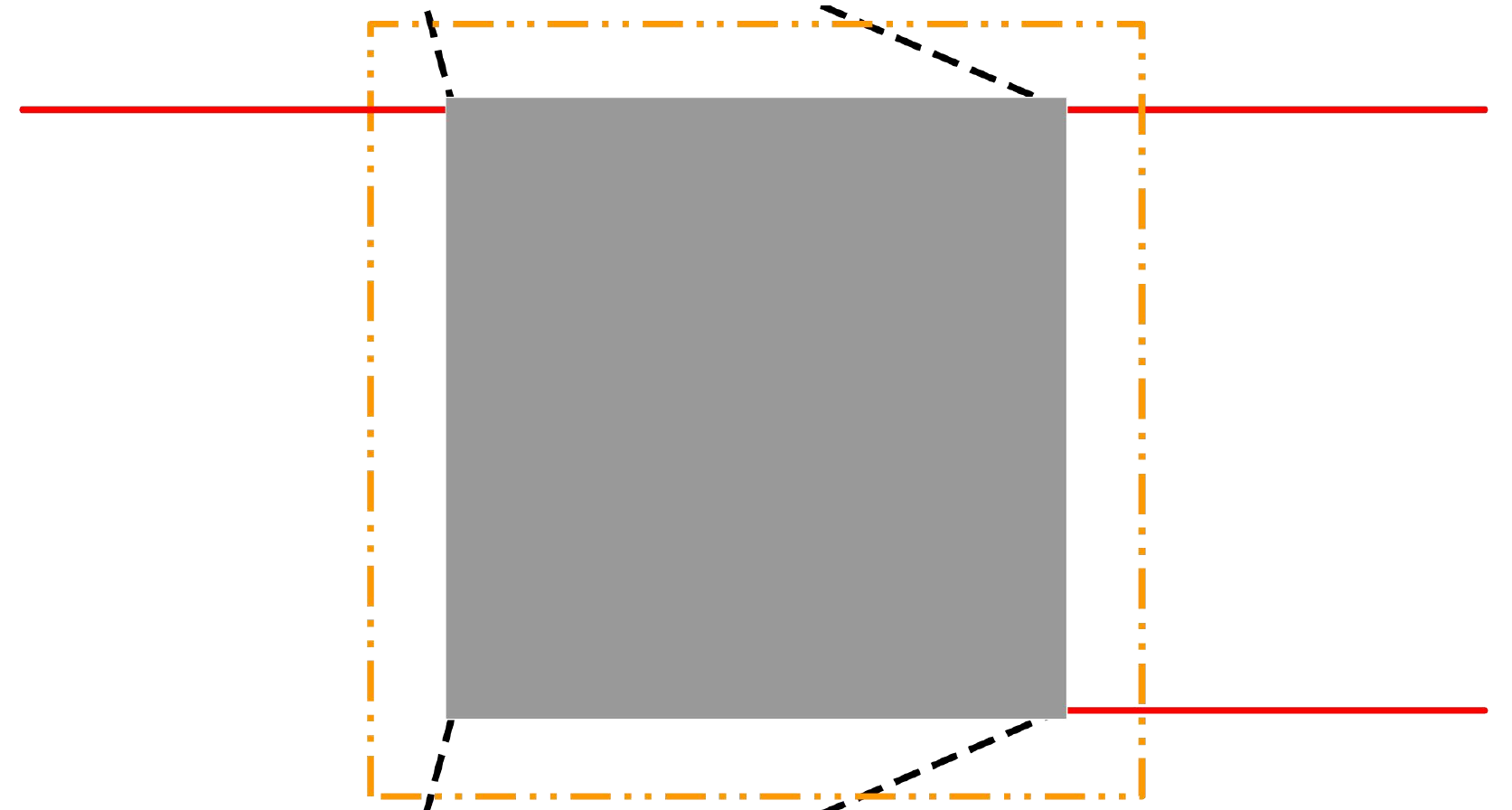}
    \caption{Dotted-dashed boxes in \cref{fig:correlator} featuring $\omega_\uGW (\eta_\uin,\bx_i,\bq_i)$ at various orders in $\zeta_{gL}$. The shadow squares should be replaced with one of the panels of Figure~3 in Ref.~\cite{Li:2023xtl}.  }\label{fig:FD_bl_Frame}
\end{figure}

As indicated by \cref{fig:FD_bl_Frame}, the dotted-dashed boxes can be used for representations of $\omega_\uGW (\eta_\uin,\bx_i,\bq_i)$ at different orders in $\zeta_{gL}$. 
From left to right, the first box represents $\omega_\uGW (\eta_\uin,\bx_i,\bq_i)$ at $\cO (\zeta_{gL})$ order, the second and third boxes represent $\omega_\uGW (\eta_\uin,\bx_i,\bq_i)$ at $\cO (\zeta_{gL}^2)$ order and the last two panels represents $\omega_\uGW (\eta_\uin,\bx_i,\bq_i)$ at $\cO (\zeta_{gL}^3)$ order. 
The shadow squares cover the connections of blue solid lines which correspond to the correlators between $\zeta_{gS}$ and $\zeta_{S}$ in Eq.~\eqref{eq:ogwexpand}. 
They should be replaced with specific panels of Figure~3 in Ref.~\cite{Li:2023xtl}, based on the count of Gaussian-vertices, $\fnl$-vertices, and $\gnl$-vertices in each panel. 
To be specific, for $\bar{\Omega}_\uGW^{(a,b)}$, we can straightforwardly express the count of $\fnl$-vertices as $N_{\fnl}^{(a,b)} = 2a$ and the count of $\gnl$-vertices as $N_{\gnl}^{(a,b)} = b$. 
As $\bar{\Omega}_\uGW \sim \langle\zeta^4\rangle$ indicates that four vertices are involved in the propagators of $\zeta$ in total, the count of Gaussian-vertices is $N_\mathrm{Gau}^{(a,b)} = 4 - 2a - b$.
We will provide a detailed explanation of the conditions that the panels replacing these shadow squares need to satisfy in the following.

\begin{figure}[htbp]
    \centering
    \includegraphics[width =0.3 \columnwidth]{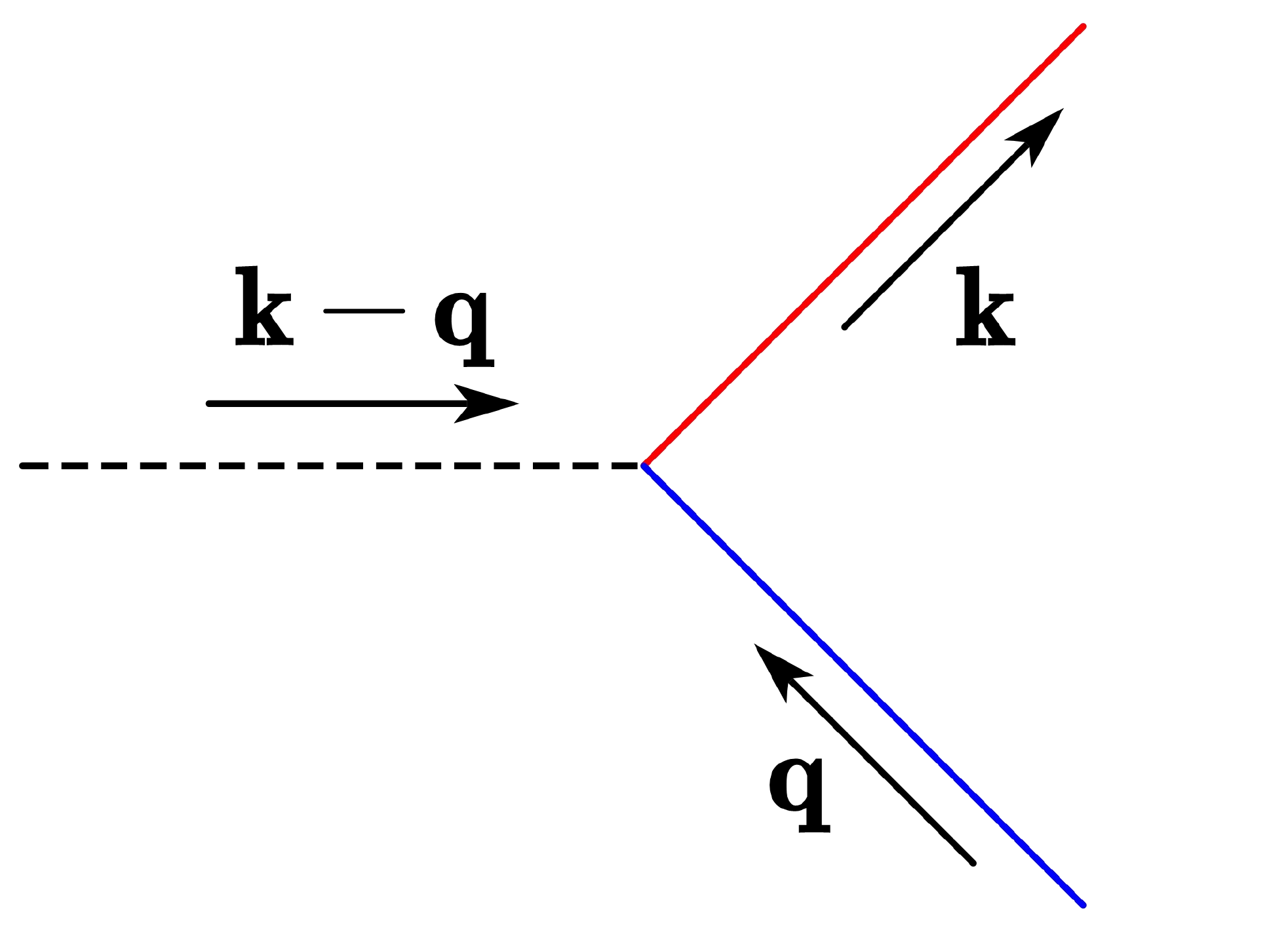}
    \hfil
    \includegraphics[width =0.3 \columnwidth]{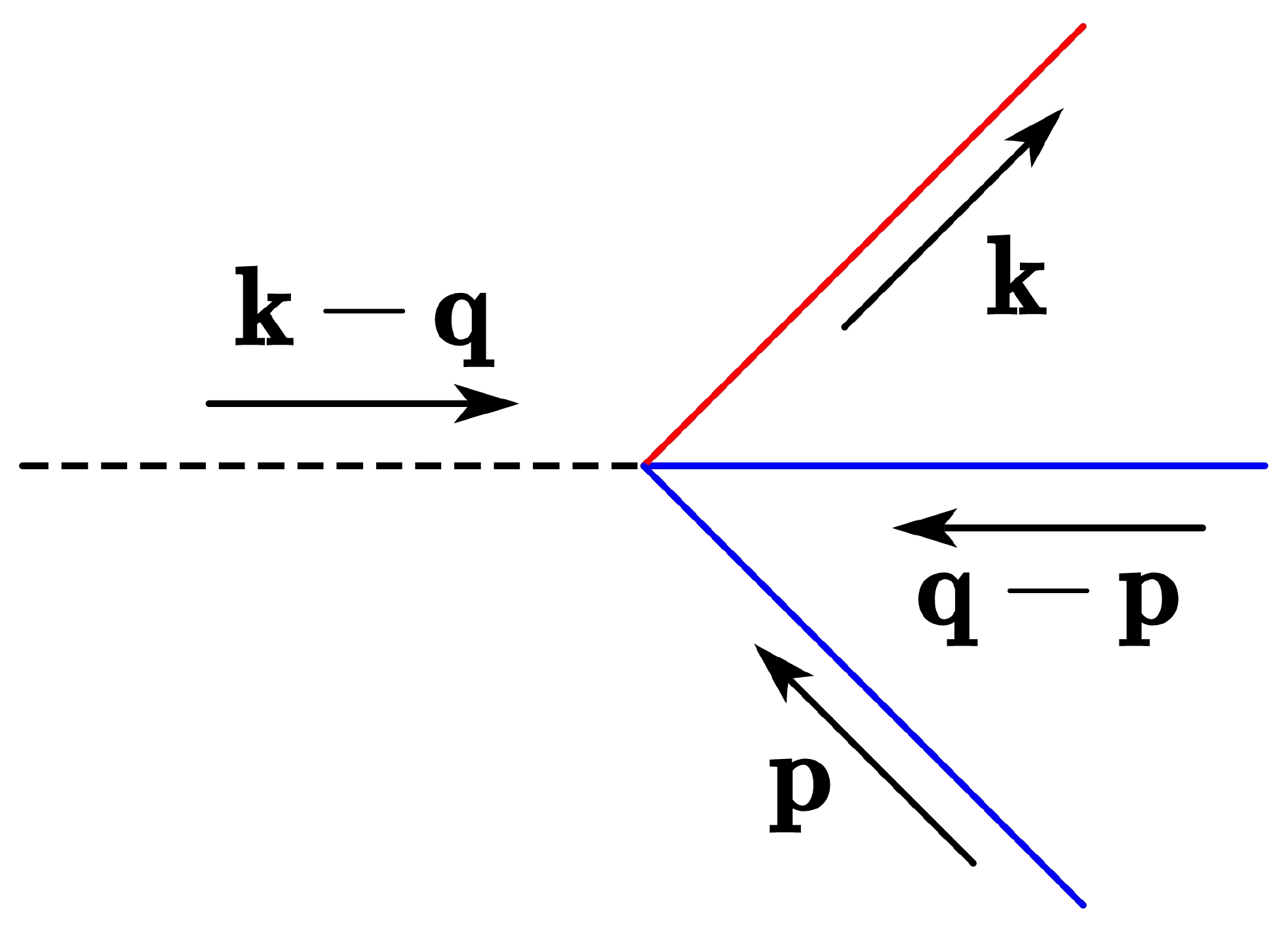}
    \hfil
    \includegraphics[width =0.3 \columnwidth]{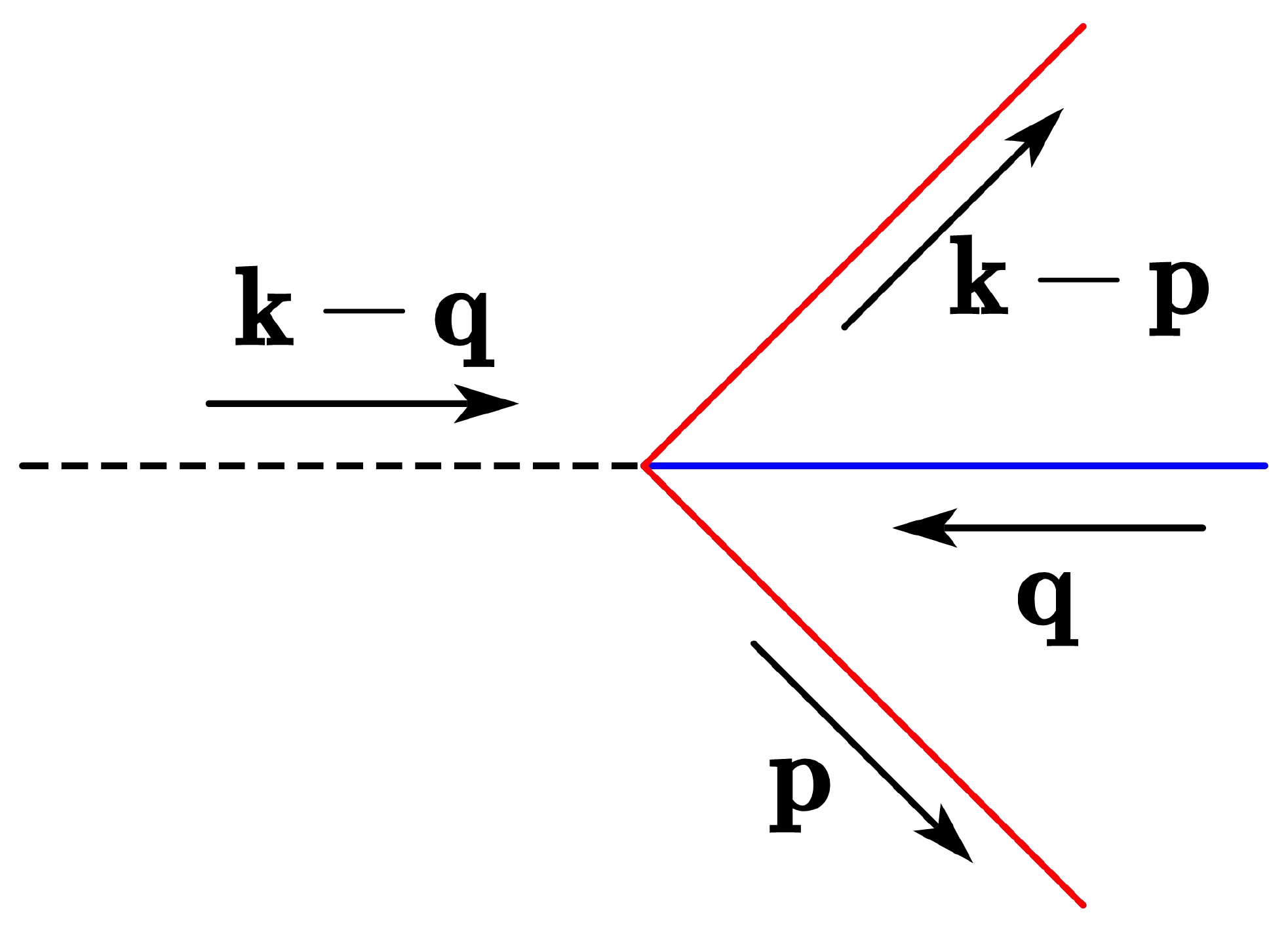}
    \caption{Vertices involved in the non-Gaussian bridges. 
    }\label{fig:v-transform}
\end{figure}

In fact, a simple approach to obtain the explicit expression of Eq.~\eqref{eq:ogwexpand} is to identify all allowable connections among the three (four) $\omega_\uGW (\eta_\uin,\bk_i,\bq_i)$ via three (four) ``non-Gaussian bridges''.
Compared with the vertices before connection, the inclusion of the ``non-Gaussian bridges'' results in the transformation of a vertex into another vertex that represents non-Gaussian parameters of higher order, as illustrated in \cref{fig:v-transform}. 
For contractions at $\cO (\zeta_{gL})$ order that correspond to the first panel of \cref{fig:FD_bl_Frame}, there are two allowable transformations: 
\begin{itemize}
    \item As depicted in the left panel of \cref{fig:v-transform}, one Gaussian-vertex is transformed into an $\fnl$-vertex, which also doubles the symmetric factor. 
    \item As depicted in the middle panel of \cref{fig:v-transform}, one $\fnl$-vertex is transformed into a $\gnl$-vertex, which triples the symmetric factor. 
\end{itemize}
Similarly, for contractions at $\cO (\zeta_{gL}^2)$ order that correspond to the second and third panels of \cref{fig:FD_bl_Frame}, there are four allowable transformations: 
\begin{itemize}
    \item Two Gaussian-vertices are transformed into $\fnl$-vertices, respectively. 
    \item Two $\fnl$-vertices are transformed into $\gnl$-vertices, respectively. 
    \item One Gaussian-vertex is transformed into an $\fnl$-vertex and one $\fnl$-vertex is transformed into a $\gnl$-vertex.
    \item As depicted in the right panel of \cref{fig:v-transform}, one Gaussian-vertex is transformed into a $\gnl$-vertex, which sextuples the symmetric factor.
\end{itemize}
For contractions at $\cO (\zeta_{gL}^3)$ order that correspond to the last two panels of \cref{fig:FD_bl_Frame}, there are six allowable transformations: 
\begin{itemize}
    \item Three Gaussian-vertices are transformed into $\fnl$-vertices, respectively. 
    \item Three $\fnl$-vertices are transformed into $\gnl$-vertices, respectively. 
    \item Two Gaussian-vertices are transformed into $\fnl$-vertices and one $\fnl$-vertex is transformed into a $\gnl$-vertex.
    \item One Gaussian-vertex is transformed into an $\fnl$-vertex and two $\fnl$-vertex are transformed into a $\gnl$-vertices.
    \item One Gaussian-vertex is transformed into an $\fnl$-vertex and another Gaussian-vertex is transformed into a $\gnl$-vertex.
    \item One Gaussian-vertex is transformed into a $\gnl$-vertex and one $\fnl$-vertex is transformed into a $\gnl$-vertex.
\end{itemize}
Accordingly, only the panels of Figure~3 in Ref.~\cite{Li:2023xtl} that contain a sufficient count of Gaussian-vertices and $\fnl$-vertices can replace the shadow squares in the corresponding boxes.  

In summary, when considering the correlation functions of $\omega_{\uGW}(\eta_\uin,\bx,\bq)$, the initial energy-density full spectrum in Eq.~(\ref{eq:ogwexpand}) can be expressed in the form of 
\begin{eqnarray}\label{eq:omega-result}
    \omega_{\uGW}(\eta_\uin,\bx,\bq) 
    &=& \frac{\bar{\Omega}_{\uGW}(\eta_\uin,q)}{4\pi} + \frac{\Omega_{\mathrm{ng}}^{(1)}(\eta_\uin,q)}{4\pi} \int \frac{\ud^{3}\bk}{(2\pi)^{3/2}} e^{i\bk\cdot\bx} \zeta_{gL}(\bk)\nonumber\\
    && + \frac{\Omega_{\mathrm{ng}}^{(2)}(\eta_\uin,q)}{4\pi} \int \frac{\ud^{3}\bk\,\ud^{3}\bp}{(2\pi)^{3}} e^{i\bk\cdot\bx} \zeta_{gL}(\bp) \zeta_{gL}(\bk - \bp) \nonumber\\
    && + \frac{\Omega_{\mathrm{ng}}^{(3)}(\eta_\uin,q)}{4\pi} \int \frac{\ud^{3}\bk\,\ud^{3}\bp\,\ud^{3}\bl}{(2\pi)^{9/2}} e^{i\bk\cdot\bx} \zeta_{gL}(\bl) \zeta_{gL}(\bp-\bl) \zeta_{gL}(\bk - \bp)\ ,
\end{eqnarray}
where $\Omega_\ung^{(1)} (\eta_\uin,q)$, corresponding to the first panel of \cref{fig:FD_bl_Frame}, is identical to that in our existing work \cite{Li:2023xtl}, i.e.,
\begin{eqnarray}\label{eq:Ong1-def}
    \Omega_\ung^{(1)} (\eta_\uin,q)
    &=& \sum_{(a,b)} \Biggl[
        \left(2\times\frac{3\fnl/5}{1}\right) \binom{N_\mathrm{Gau}^{(a,b)}}{1} 
        + \left(3\times\frac{9\gnl/25}{3\fnl/5}\right) \binom{N_{\fnl}^{(a,b)}}{1}  
    \Biggr] \bar{\Omega}_\uGW^{(a,b)} \nonumber\\
    &=& \frac{6\fnl}{5} \left(4 \bar{\Omega}_\uGW^{(0,0)} + 3 \bar{\Omega}_\uGW^{(0,1)} + 2 \bar{\Omega}_\uGW^{(1,0)}
    + 2 \bar{\Omega}_\uGW^{(0,2)} +  \bar{\Omega}_\uGW^{(1,1)} +  \bar{\Omega}_\uGW^{(0,3)}\right)\nonumber\\
    && + \frac{9 \gnl}{5 \fnl} \left(2 \bar{\Omega}_\uGW^{(1,0)} + 2 \bar{\Omega}_\uGW^{(1,1)} + 4 \bar{\Omega}_\uGW^{(2,0)} + 2  \bar{\Omega}_\uGW^{(1,2)}\right)\ ,
\end{eqnarray} 
$\Omega_\ung^{(2)} (\eta_\uin,q)$, corresponding to the second and third panels of \cref{fig:FD_bl_Frame}, is derived for the first time, given by 
\begin{eqnarray}\label{eq:Ong2-def}
    \Omega_\ung^{(2)} (\eta_\uin,q) 
    &=& \sum_{(a,b)} \Biggl[
        \left(2\times\frac{3\fnl/5}{1}\right)^2 \binom{N_\mathrm{Gau}^{(a,b)}}{2} 
        + \left(3\times\frac{9\gnl/25}{3\fnl/5}\right)^2 \binom{N_{\fnl}^{(a,b)}}{2} \nonumber\\
        &&\qquad\ + \left(2\times\frac{3\fnl/5}{1}\right) \left(3\times\frac{9\gnl/25}{3\fnl/5}\right) \binom{N_\mathrm{Gau}^{(a,b)}}{1} \binom{N_{\fnl}^{(a,b)}}{1}  \nonumber\\
        &&\qquad\ + \left(6\times\frac{9\gnl/25}{1}\right) \binom{N_\mathrm{Gau}^{(a,b)}}{1} 
    \Biggr] \bar{\Omega}_\uGW^{(a,b)} \nonumber\\
    &=& \left(\frac{6\fnl}{5}\right)^2 \left(6 \bar{\Omega}_\uGW^{(0,0)} + 3 \bar{\Omega}_\uGW^{(0,1)} + \bar{\Omega}_\uGW^{(1,0)} +  \bar{\Omega}_\uGW^{(0,2)}\right)\nonumber\\
    && + \left(\frac{9\gnl}{5\fnl}\right)^2 \left(\bar{\Omega}_\uGW^{(1,0)} + \bar{\Omega}_\uGW^{(1,1)} + 6 \bar{\Omega}_\uGW^{(2,0)} + \bar{\Omega}_\uGW^{(1,2)}\right)\nonumber\\
    && + \frac{54 \gnl}{25} \left(4 \bar{\Omega}_\uGW^{(0,0)} + 3 \bar{\Omega}_\uGW^{(0,1)}  + 6 \bar{\Omega}_\uGW^{(1,0)}+ 2 \bar{\Omega}_\uGW^{(0,2)} + 3 \bar{\Omega}_\uGW^{(1,1)}
    + \bar{\Omega}_\uGW^{(0,3)}\right)\ .
\end{eqnarray} 
and $\Omega_\ung^{(3)} (\eta_\uin,q)$, corresponding to the last two panels of \cref{fig:FD_bl_Frame}, is also derived for the first time, given by 
\begin{eqnarray}\label{eq:Ong3-def}
    \Omega_\ung^{(3)} (\eta_\uin,q) 
    &=& \sum_{(a,b)} \Biggl[
        \left(2\times\frac{3\fnl/5}{1}\right)^3 \binom{N_\mathrm{Gau}^{(a,b)}}{3} + \left(3\times\frac{9\gnl/25}{3\fnl/5}\right)^3 \binom{N_{\fnl}^{(a,b)}}{3} \nonumber\\
        &&\qquad\ + \left(2\times\frac{3\fnl/5}{1}\right)^2 \left(3\times\frac{9\gnl/25}{3\fnl/5}\right) \binom{N_\mathrm{Gau}^{(a,b)}}{2} \binom{N_{\fnl}^{(a,b)}}{1} \nonumber\\
        &&\qquad\ + \left(2\times\frac{3\fnl/5}{1}\right) \left(3\times\frac{9\gnl/25}{3\fnl/5}\right)^2 \binom{N_\mathrm{Gau}^{(a,b)}}{1} \binom{N_{\fnl}^{(a,b)}}{2} \nonumber\\
        &&\qquad\ + \left(6\times\frac{9\gnl/25}{1}\right) \left(2\times\frac{3\fnl/5}{1}\right) N_\mathrm{Gau}^{(a,b)} \left(N_\mathrm{Gau}^{(a,b)} - 1\right) \nonumber\\
        &&\qquad\ + \left(6\times\frac{9\gnl/25}{1}\right) \left(3\times\frac{9\gnl/25}{3\fnl/5}\right) \binom{N_\mathrm{Gau}^{(a,b)}}{1} \binom{N_{\fnl}^{(a,b)}}{1} 
    \Biggr] \bar{\Omega}_\uGW^{(a,b)} \nonumber\\
    &=& \left(\frac{6\fnl}{5}\right)^3 \left(4 \bar{\Omega}_\uGW^{(0,0)} + \bar{\Omega}_\uGW^{(0,1)}\right) \nonumber\\
    && + 3\times\left(\frac{6}{5}\right)^3 \fnl\gnl \left(6 \bar{\Omega}_\uGW^{(0,0)} + 3 \bar{\Omega}_\uGW^{(0,1)} + 2 \bar{\Omega}_\uGW^{(1,0)} + \bar{\Omega}_\uGW^{(0,2)}\right) \nonumber\\
    && + 2\times\left(\frac{9}{5}\right)^3 \frac{\gnl^2}{\fnl} \left(2 \bar{\Omega}_\uGW^{(1,0)} + \bar{\Omega}_\uGW^{(1,1)}\right) + 4\times\left(\frac{9\gnl}{5\fnl}\right)^3 \bar{\Omega}_\uGW^{(2,0)} \ .
\end{eqnarray} 
Notably, despite the presence of $\fnl$ in denominators in Eqs.~(\ref{eq:Ong1-def},\ref{eq:Ong2-def},\ref{eq:Ong3-def}), the value of $\fnl$ could be taken as zero, since $\bar{\Omega}_\uGW^{(a,b)}$ is proportional to $\fnl^{2a}$ and thus $\fnl$ disappears from the denominators after reduction. 
It is further observed that $\Omega_\ung^{(3)}$ becomes zero when $\fnl=0$.

Therefore, the present-day density contrast of \acp{SIGW}, as defined in Eq.~\eqref{eq:delta-0}, can be decomposed into three terms according to their orders in $\zeta_{gL}$, namely  
\begin{equation}\label{eq:delta-0-12}
    \delta_{\uGW,0}(\bq) = \delta_{\uGW,0}^{(1)}(\bq) + \delta_{\uGW,0}^{(2)}(\bq) + \delta_{\uGW,0}^{(3)}(\bq)\ ,
\end{equation}
where we introduce three quantities of the form 
\begin{eqnarray}
    \delta_{\uGW,0}^{(1)}(\bq) 
        &=& \biggl\{
            \frac{\Omega_{\mathrm{ng}}^{(1)} (\eta_\uin,2\pi\nu)}{\bar{\Omega}_\uGW (\eta_\uin,2\pi\nu)}
            + \frac{3}{5} \bigl[4 - n_{\uGW} (\nu)\bigr]
        \biggr\}
        \int \frac{\ud^{3}\bk}{(2\pi)^{3/2}} e^{i\bk\cdot\bx} \zeta_{gL}(\bk)\ ,\label{eq:delta-0-1}\\
    \delta_{\uGW,0}^{(2)}(\bq) 
        &=& \biggl\{
            \frac{\Omega_{\mathrm{ng}}^{(2)} (\eta_\uin,2\pi\nu)}{\bar{\Omega}_\uGW (\eta_\uin,2\pi\nu)}
            + \frac{9}{25} \fnl \bigl[4 - n_{\uGW} (\nu)\bigr]
        \biggr\}
        \int \frac{\ud^{3}\bk\,\ud^{3}\bp}{(2\pi)^{3}} e^{i\bk\cdot\bx} \zeta_{gL}(\bp) \zeta_{gL}(\bk - \bp)\ ,\label{eq:delta-0-2}\\
    \delta_{\uGW,0}^{(3)}(\bq) 
        &=& \biggl\{
            \frac{\Omega_{\mathrm{ng}}^{(3)} (\eta_\uin,2\pi\nu)}{\bar{\Omega}_\uGW (\eta_\uin,2\pi\nu)}
            + \frac{27}{125} \gnl \bigl[4 - n_{\uGW} (\nu)\bigr]
        \biggr\}
        \int \frac{\ud^{3}\bk\,\ud^{3}\bp\,\ud^{3}\bl}{(2\pi)^{9/2}} e^{i\bk\cdot\bx} \zeta_{gL}(\bl) \zeta_{gL}(\bp - \bl) \zeta_{gL}(\bk - \bp)\ .\nonumber\\
        && \label{eq:delta-0-3}
\end{eqnarray}
Here, as the long-wavelength mode reentered the Hubble horizon during matter domination, we have also considered the Bardeen potential up to third order in $\zeta_{gL}$. 
It is now given by 
\begin{eqnarray}\label{eq:Phi}
    \Phi (\eta_\uin,\bx) 
    = \frac{3}{5} \int \frac{\ud^{3}\bk}{(2\pi)^{3/2}} e^{i\bk\cdot\bx} 
    &\biggl[&
        \zeta_{gL}(\bk) + \frac{3}{5}\fnl \int \frac{\ud^{3}\bp}{(2\pi)^{3/2}} \zeta_{gL}(\bp) \zeta_{gL}(\bk - \bp)\\ 
        &&+ \frac{9}{25} \gnl \int \frac{\ud^{3}\bp\,\ud^{3}\bl}{(2\pi)^{3}} e^{i\bk\cdot\bx} \zeta_{gL}(\bl) \zeta_{gL}(\bp - \bl) \zeta_{gL}(\bk - \bp)
    \biggr]\ .\nonumber
\end{eqnarray}

\subsection{Angular bispectrum}
For $\tilde{b}_{\ell_1 \ell_2 \ell_3} (\nu)$ in Eq.~\eqref{eq:btilde-def}, an integral is involved due to the three-point correlation function of $\delta_{\uGW,0}$ in $\ell,m$ space. 
Non-vanishing contractions at leading order in $A_L$ comprise two $\delta_{\uGW,0}^{(1)}$ and one $\delta_{\uGW,0}^{(2)}$ since $\zeta_{gL}$ is Gaussian. 
According to the Wick's theorem, the four-point correlator of $\zeta_{gL}$ therein leads to contractions at $\cO (A_L^2)$ order with a Dirac $\delta$ function. 
Further, we use the line-of-sight relation $\bx-\bx' = (\eta_\uin-\eta_0)(\bn_0-\bn_0')$ and the identity $e^{i k \mu (\eta_\uin-\eta_0)} = 4 \pi \sum_{\ell m}  (-i)^\ell j_\ell [k (\eta_0-\eta_\uin)] Y^\ast_{\ell m} (\hat{\bk}) Y_{\ell m} (\bn_0)$ for subsequent calculations. 
By utilizing the representation of the Dirac $\delta$ function in terms of spherical harmonics and the orthonormality of spherical harmonics, the integral for $\tilde{b}_{\ell_1 \ell_2 \ell_3} (\nu)$ becomes 
\begin{eqnarray}\label{eq:b-int}
    \tilde{b}_{\ell_1 \ell_2 \ell_3} (\nu)
    &=& \int_0^\infty \ud r\, r^2 
    \prod_{i=1}^3
    \left[
        \frac{2}{\pi} \int_0^\infty \ud k_i\, k_i^2 j_{\ell_i} (k_i (\eta_0 - \eta_\uin)) j_{\ell_i} (k_i r) 
    \right]
    B(2\pi\nu,k_1,k_2,k_3) \ ,
\end{eqnarray}
where $B(q,k_1,k_2,k_3)$ is introduced by the three-point correlation of density contrast in Fourier mode, i.e., 
\begin{eqnarray}
    B(q,k_1,k_2,k_3) &=& 2 \biggl\{
            \frac{\Omega_{\mathrm{ng}}^{(1)} (\eta_\uin,q)}{\bar{\Omega}_\uGW (\eta_\uin,q)}
            + \frac{3}{5} \biggl[4 - n_{\uGW} \left(\frac{q}{2\pi}\right)\biggr]
        \biggr\}^2 
        \biggl\{
            \frac{\Omega_{\mathrm{ng}}^{(2)} (\eta_\uin,q)}{\bar{\Omega}_\uGW (\eta_\uin,q)}
            + \frac{9}{25} \fnl \biggl[4 - n_{\uGW} \left(\frac{q}{2\pi}\right)\biggr]
        \biggr\}\nonumber\\
        && \times\left[
            \frac{2\pi^2}{k_1^3} \Delta_{L}^2 (k_1) \frac{2\pi^2}{k_2^3} \Delta_{L}^2 (k_2) 
            + \frac{2\pi^2}{k_2^3} \Delta_{L}^2 (k_2) \frac{2\pi^2}{k_3^3} \Delta_{L}^2 (k_3)
            + \frac{2\pi^2}{k_3^3} \Delta_{L}^2 (k_3) \frac{2\pi^2}{k_1^3} \Delta_{L}^2 (k_1)
        \right] \ .\nonumber\\
        &&
\end{eqnarray}
For simplicity, we set $\Delta_{L}^{2} \simeq A_L$ in the following derivation. 
Further, by utilizing the closure relation of Bessel function 
\begin{eqnarray}
    &&\frac{2}{\pi} \int_0^\infty \ud k\, k^2 j_{\ell} [k (\eta_0 - \eta_\uin)] j_{\ell} (k r) = \frac{\delta (\eta_0 - \eta_\uin - r)}{r^2}\ ,
\end{eqnarray}
and the integral of the form
\begin{equation}
\int_0^\infty \ud \ln k\, j_{\ell}^2 (k \left(\eta_0 - \eta_\uin\right)) = \frac{1}{2 \ell (\ell + 1)} \ , 
\end{equation}
we can complete the integration of Eq.~\eqref{eq:b-int} and then get one of our leading results as follows 
\begin{equation}\label{eq:btilde-res}
    \tilde{b}_{\ell_1 \ell_2 \ell_3} (\nu)
    = b (\nu) \left[
        \frac{1}{\ell_1 \ell_2 \left(\ell_1 + 1\right)\left(\ell_2 + 1\right)} + \frac{1}{\ell_2 \ell_3 \left(\ell_2 + 1\right)\left(\ell_3 + 1\right)} + \frac{1}{\ell_3 \ell_1 \left(\ell_3 + 1\right)\left(\ell_1 + 1\right)}
    \right]\ ,
\end{equation}
where we introduce a function $b (\nu)$ as follows
\begin{equation}\label{eq:b-def}
    b (\nu) = 8\pi^2 A_{L}^2 
    \biggl\{
        \frac{\Omega_{\mathrm{ng}}^{(1)} (\eta_\uin,2\pi\nu)}{\bar{\Omega}_\uGW (\eta_\uin,2\pi\nu)}
        + \frac{3}{5} \bigl[4 - n_{\uGW} \left(\nu\right)\bigr]
    \biggr\}^2
    \biggl\{
        \frac{\Omega_{\mathrm{ng}}^{(2)} (\eta_\uin,2\pi\nu)}{\bar{\Omega}_\uGW (\eta_\uin,2\pi\nu)}
        + \frac{9}{25} \fnl \bigl[4 - n_{\uGW} \left(\nu\right)\bigr]
    \biggr\}\ .
\end{equation}
It is worth noting that Eqs.~(\ref{eq:Ong1-def},\ref{eq:Ong2-def},\ref{eq:b-def}) imply that $b(\nu) = 0$ when $\fnl = \gnl = 0$. 
In other words, the Gaussian $\zeta$ always leads to a Gaussian background of \acp{SIGW}, which is in line with our expectation.

\subsection{Angular trispectrum}
Analogous to \ac{CMB} \cite{Kogo:2006kh}, the reduced angular trispectrum $t^{\ell_1 \ell_2}_{\ell_3 \ell_4} (L,\nu)$ for \acp{SIGW}, as defined in Eqs.~(\ref{eq:Tl-def},\ref{eq:Pl-def},\ref{eq:tl-def}), can be obtained after a lengthy but similar derivation. 
As depicted in \cref{fig:correlator}, the non-vanishing connected contractions at the leading order in $A_L$ correspond to two types of Feynman-like diagrams. 
One consists of two $\delta_{\uGW,0}^{(2)}$ and two $\delta_{\uGW,0}^{(1)}$ corresponding to the middle panel of \cref{fig:correlator}, while the other consists of one $\delta_{\uGW,0}^{(3)}$ and three $\delta_{\uGW,0}^{(1)}$ corresponding to the right panel of \cref{fig:correlator}.
According to the permutation symmetry, $t^{\ell_1 \ell_2}_{\ell_3 \ell_4} (L,\nu)$ exactly corresponds to 
\begin{eqnarray}
    && \frac{1}{2} \left\langle\delta_{\uGW,0}^{(2)} (\bq_1) \delta_{\uGW,0}^{(1)} (\bq_2) \delta_{\uGW,0}^{(2)} (\bq_3) \delta_{\uGW,0}^{(1)} (\bq_4)\right\rangle_c \nonumber\\ 
    && + \frac{1}{6} \Bigl[ \left\langle\delta_{\uGW,0}^{(3)} (\bq_1) \delta_{\uGW,0}^{(1)} (\bq_2) \delta_{\uGW,0}^{(1)} (\bq_3) \delta_{\uGW,0}^{(1)} (\bq_4)\right\rangle_c 
    + \left\langle\delta_{\uGW,0}^{(1)} (\bq_1) \delta_{\uGW,0}^{(1)} (\bq_2) \delta_{\uGW,0}^{(3)} (\bq_3) \delta_{\uGW,0}^{(1)} (\bq_4)\right\rangle_c \Bigr]\ ,\nonumber
\end{eqnarray}
where the subscript $_c$ denotes the connected part of the four-point correlators. 
The combination of these correlators in $\ell,m$ space can lead to 
\begin{eqnarray}\label{eq:tl-int}
    t^{\ell_1 \ell_2}_{\ell_3 \ell_4} (L,\nu) 
    &=& \Biggl\{
        4 \biggl\{
            \frac{\Omega_{\mathrm{ng}}^{(1)} (\eta_\uin,2\pi\nu)}{\bar{\Omega}_\uGW (\eta_\uin,2\pi\nu)}
            + \frac{3}{5} \bigl[4 - n_{\uGW} (\nu)\bigr]
        \biggr\}^2 
        \biggl\{
            \frac{\Omega_{\mathrm{ng}}^{(2)} (\eta_\uin,2\pi\nu)}{\bar{\Omega}_\uGW (\eta_\uin,2\pi\nu)}
            + \frac{9}{25} \fnl \bigl[4 - n_{\uGW} (\nu)\bigr]
        \biggr\}^2\nonumber\\
        &&\qquad \times \int \ud r_1\, r_1^2 \, \ud r_2\, r_2^2\, F_L (r_1,r_2) 
            \alpha_{\ell_1} (r_1) \beta_{\ell_2} (r_1) \alpha_{\ell_3} (r_2) \beta_{\ell_4} (r_2) \nonumber\\
        &&\quad + \frac{1}{4} \biggl\{
            \frac{\Omega_{\mathrm{ng}}^{(1)} (\eta_\uin,2\pi\nu)}{\bar{\Omega}_\uGW (\eta_\uin,2\pi\nu)}
            + \frac{3}{5} \bigl[4 - n_{\uGW} (\nu)\bigr]
        \biggr\}^3 
        \biggl\{
            \frac{\Omega_{\mathrm{ng}}^{(3)} (\eta_\uin,2\pi\nu)}{\bar{\Omega}_\uGW (\eta_\uin,2\pi\nu)}
            + \frac{27}{125} \gnl \bigl[4 - n_{\uGW} (\nu)\bigr]
        \biggr\}\nonumber\\
        &&\qquad \times \int r^2 \ud r\, 
        \bigl[\alpha_{\ell_1} (r) \beta_{\ell_3} (r) + \beta_{\ell_1} (r) \alpha_{\ell_3} (r)\bigr] \beta_{\ell_2} (r) \beta_{\ell_4} (r) 
    \Biggr\}
    h_{\ell_1 \ell_2 L} h_{\ell_3 \ell_4 L} \ ,
\end{eqnarray}
where $h_{\ell\ell'\ell''}$ has been defined in Eq.~(\ref{eq:hlll}), and we further introduce three integrals of the form 
\begin{eqnarray}
&& F_L (r_1,r_2) = 4 \pi \int \frac{\ud K}{K}\, \Delta_{gL} (K) j_{L} (K r_1) j_{L} (K r_2)\ ,\\
&& \alpha_\ell (r) = \frac{2}{\pi} \int \ud k\, k^2 j_{\ell} (k r) j_{\ell} [k (\eta_0-\eta_\uin)]\ ,\\
&& \beta_\ell (r) = 4 \pi \int \frac{\ud k}{k}\, \Delta_{gL} (k) j_{\ell} (k r) j_{\ell} [k (\eta_0-\eta_\uin)]\ .
\end{eqnarray}
In particular, for the scale-invariant $\Delta_{gL} (k) \simeq A_L$ considered in this work, we reduce $t^{\ell_1 \ell_2}_{\ell_3 \ell_4} (L,\nu)$ to 
\begin{equation}\label{eq:t-res}
    t^{\ell_1 \ell_2}_{\ell_3 \ell_4} (L,\nu) 
    = \frac{1}{\ell_2 (\ell_2 + 1)} \frac{1}{\ell_4 (\ell_4 + 1)} 
    \Biggl[
        \frac{t_1 (\nu)}{L (L + 1)} + \frac{t_2 (\nu)}{\ell_1 (\ell_1 + 1)} + \frac{t_2 (\nu)}{\ell_3 (\ell_3 + 1)} 
    \Biggr] 
    h_{\ell_1 \ell_2 L} h_{\ell_3 \ell_4 L} \ ,
\end{equation}
where $t_1 (\nu)$ and $t_2 (\nu)$ are introduced as follows 
\begin{eqnarray}
    t_1 (\nu) &=& 4 \times \bigl(2 \pi A_L\bigr)^3 
        \biggl\{
            \frac{\Omega_{\mathrm{ng}}^{(1)} (\eta_\uin,2\pi\nu)}{\bar{\Omega}_\uGW (\eta_\uin,2\pi\nu)}
            + \frac{3}{5} \bigl[4 - n_{\uGW} (\nu)\bigr]
        \biggr\}^2 
        \biggl\{
            \frac{\Omega_{\mathrm{ng}}^{(2)} (\eta_\uin,2\pi\nu)}{\bar{\Omega}_\uGW (\eta_\uin,2\pi\nu)}
            + \frac{9}{25} \fnl \bigl[4 - n_{\uGW} (\nu)\bigr]
        \biggr\}^2 \ ,\nonumber\\
        &&\label{eq:t1-def}\\
    t_2 (\nu) &=& \frac{\bigl(2 \pi A_L\bigr)^3}{4}
        \biggl\{
            \frac{\Omega_{\mathrm{ng}}^{(1)} (\eta_\uin,2\pi\nu)}{\bar{\Omega}_\uGW (\eta_\uin,2\pi\nu)}
            + \frac{3}{5} \bigl[4 - n_{\uGW} (\nu)\bigr]
        \biggr\}^3 
        \biggl\{
            \frac{\Omega_{\mathrm{ng}}^{(3)} (\eta_\uin,2\pi\nu)}{\bar{\Omega}_\uGW (\eta_\uin,2\pi\nu)}
            + \frac{27}{125} \gnl \bigl[4 - n_{\uGW} (\nu)\bigr]
        \biggr\} \ .\nonumber\\
        &&\label{eq:t2-def}
\end{eqnarray}
It also stands for one of our leading results.

\subsection{Frequency and multipole dependence}

To investigate the frequency and multipole dependence for the angular bispectrum and trispectrum, we can reformulate $\tilde{b}_{\ell_1 \ell_2 \ell_3} (\nu)$ and $t^{\ell_1 \ell_2}_{\ell_3 \ell_4} (L,\nu)$ in terms of $\tilde{C}_\ell (\nu)$, namely 
\begin{eqnarray}
    \tilde{b}_{\ell_1 \ell_2 \ell_3} (\nu)
    &=& 2 \times 
    \biggl[
        \frac{\Omega_{\mathrm{ng}}^{(2)}}{\bar{\Omega}_\uGW}
        + \frac{9}{25} \fnl \bigl(4 - n_{\uGW}\bigr)
    \biggr]
    \biggl[
        \frac{\Omega_{\mathrm{ng}}^{(1)}}{\bar{\Omega}_\uGW}
        + \frac{3}{5} \bigl(4 - n_{\uGW}\bigr)
    \biggr]^{-2}\nonumber\\
    && \times \left(\tilde{C}_{\ell_1} \tilde{C}_{\ell_2} + \tilde{C}_{\ell_1} \tilde{C}_{\ell_3} + \tilde{C}_{\ell_2} \tilde{C}_{\ell_3}\right)\ ,\label{eq:b-Cl}\\
    t^{\ell_1 \ell_2}_{\ell_3 \ell_4} (L,\nu) 
    &=&  
    \Biggl\{
        \biggl[
            \frac{\Omega_{\mathrm{ng}}^{(3)}}{\bar{\Omega}_\uGW}
            + \frac{27}{125} \gnl \bigl(4 - n_{\uGW}\bigr)
        \biggr]
        \biggl[
            \frac{\Omega_{\mathrm{ng}}^{(1)}}{\bar{\Omega}_\uGW}
            + \frac{3}{5} \bigl(4 - n_{\uGW}\bigr)
        \biggr]^{-5} 
        \bigl(\tilde{C}_{\ell_1} + \tilde{C}_{\ell_3}\bigr)\\
        &&
        +\biggl[
            \frac{\Omega_{\mathrm{ng}}^{(2)}}{\bar{\Omega}_\uGW}
            + \frac{9}{25} \fnl \bigl(4 - n_{\uGW}\bigr)
        \biggr]^2 
        \biggl[
            \frac{\Omega_{\mathrm{ng}}^{(1)}}{\bar{\Omega}_\uGW}
            + \frac{3}{5} \bigl(4 - n_{\uGW}\bigr)
        \biggr]^{-4} \tilde{C}_{L}
    \Biggr\} 
    h_{\ell_1 \ell_2 L} h_{\ell_3 \ell_4 L} \tilde{C}_{\ell_2} \tilde{C}_{\ell_4}\ . \nonumber \label{eq:tl-Cl}
\end{eqnarray}
Here, $\tilde{C}_\ell (\nu)$ is already defined in Eq.~\eqref{eq:Ctilde-def} and its full analysis can be found in Ref.~\cite{Li:2023xtl}.
It is given by 
\begin{equation}\label{eq:reduced-APS}
    \tilde{C}_\ell (\nu) 
    = \frac{2\pi A_L}{\ell (\ell+1)}
        \left\{
            \frac{\Omega_{\mathrm{ng}}^{(1)} (\eta_\uin,2\pi\nu)}{\bar{\Omega}_\uGW (\eta_\uin,2\pi\nu)}
            + \frac{3}{5} \left[4 - n_{\uGW} (\nu)\right]
        \right\}^2\ .
\end{equation}

We find that the frequency dependence is non-trivial, since Eqs.~(\ref{eq:b-Cl},\ref{eq:tl-Cl}) are neither determined by the energy-density fraction spectrum nor the angular power spectrum. 
This implies that the angular bispectrum and trispectrum contains valuable information, which may be useful for extracting the \ac{SIGW} signal from astrophysical foregrounds. 
It is worth noting that the above results differ from the existing works. 
The frequency dependence in the study of \ac{CMB} is typically not considered due to the tightly couplings of particles in plasma, which is of thermal equilibrium in the early universe. 
As for a seminal work in Ref.~\cite{Bartolo:2019zvb}, a fraction of contributions from $\fnl$ to the angular bispectrum of \acp{SIGW} was considered, which, although frequency-dependent, can be uniquely determined by the angular power spectrum.

We also find that the multipole dependence for the angular bispectrum in Eq.~(\ref{eq:b-Cl}) is uniquely determined by that of angular power spectrum. 
In contrast, the multipole dependence for the angular trispectrum in Eq.~(\ref{eq:tl-Cl}) is determined by those of angular power spectrum and $h_{\ell\ell'\ell''}$ which has been introduced by Eq.~(\ref{eq:hlll}).

\section{Numerical results}\label{sec:num}

Based on the semi-analytic results in the former section, we can correspondingly display our numerical results in this section. 

\subsection{Angular bispectrum}\label{sec:num-b}

\begin{figure}[htbp]
    \centering
    \includegraphics[width = 1 \columnwidth]{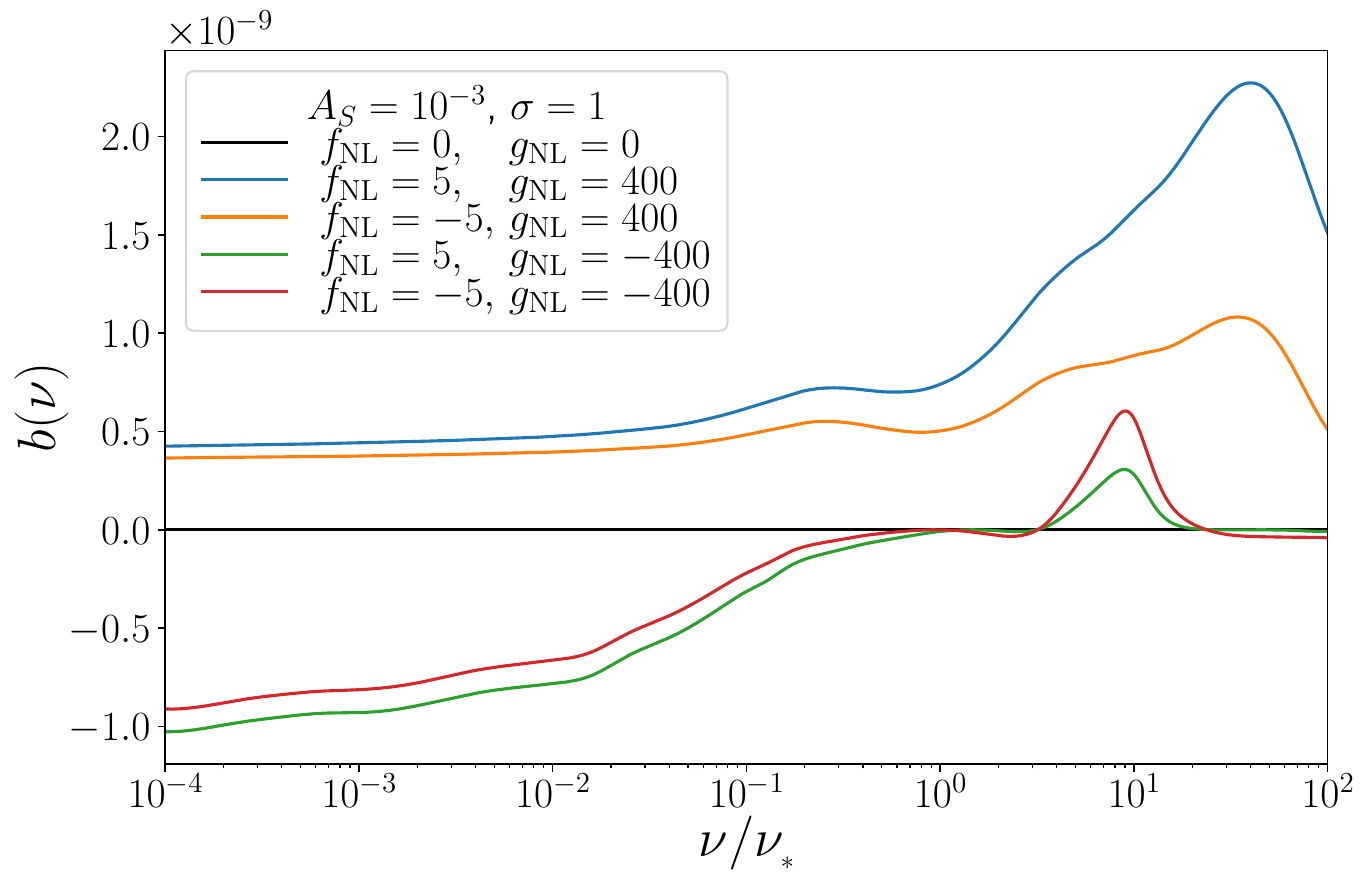}
    \caption{Frequency dependence of the reduced angular bispectrum of SIGWs.  }\label{fig:b_tilde_lin}
\end{figure}

\begin{figure*}[htbp]
    \centering
    \includegraphics[width = 1 \textwidth]{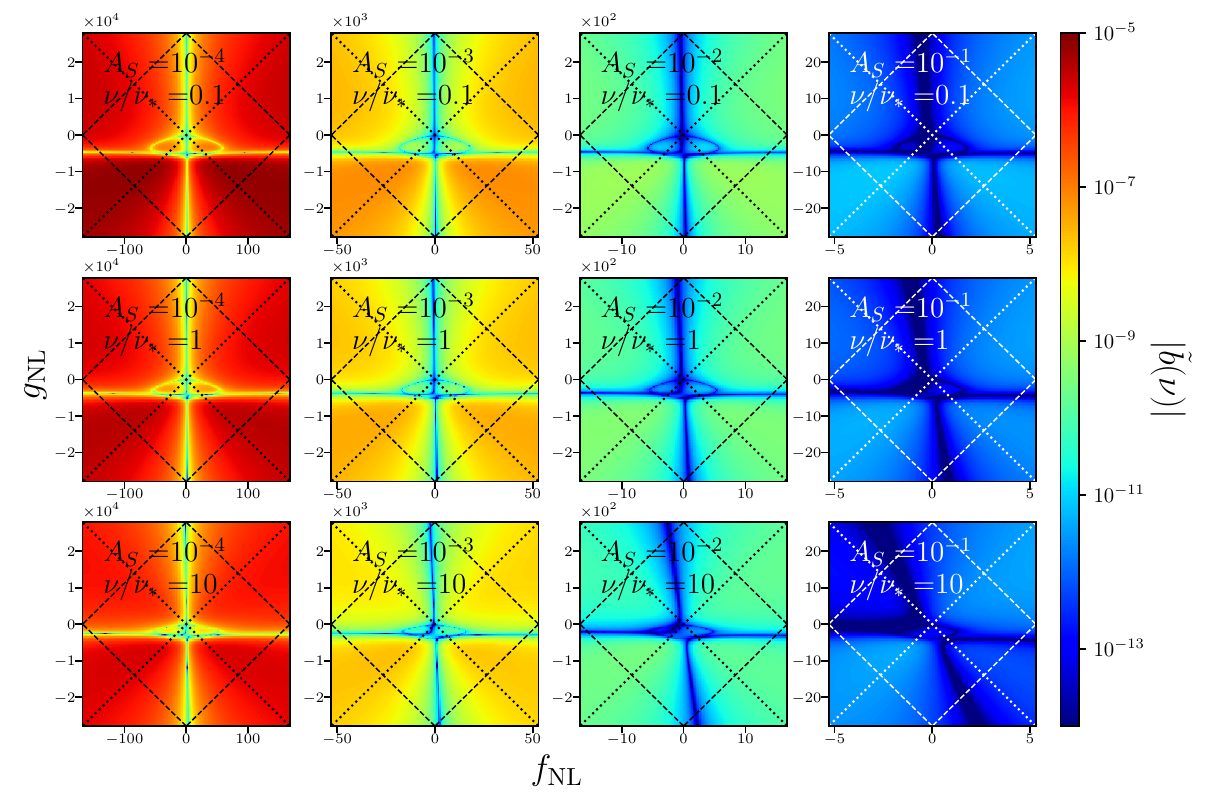}
    \caption{Frequency-dependent function for the reduced angular bispectrum of SIGWs with respect to the primordial non-Gaussian parameters $\fnl$ and $\gnl$. {  Dotted lines represent $\gnl = \pm 5 \fnl / (3 \sqrt{A_S})$, which signify that the contribution from $\fnl$ to $\zeta$ is equal to that from $\gnl$. Dashed lines represent $3|\fnl|\sqrt{A_S}/5 + 9|\gnl|A_S/25 = 1$, which signify that the contribution from non-Gaussian terms to $\zeta$ is equal to that from the Gaussian term.}
    }\label{fig:b_all}
\end{figure*}

While the multipole dependence of the reduced angular bispectrum, as shown in Eq.~\eqref{eq:btilde-res}, is straightforward, we examine the function that characterizes the frequency dependence, denoted as $b (\nu)$ in Eq.~\eqref{eq:b-def}. 
We numerically calculate $b (\nu)$ based on the numerical outcomes of $\bar{\Omega}_\uGW^{(a,b)}$ depicted in Figure~5 of Ref.~\cite{Li:2023xtl}, which is available for \acp{SIGW} produced during radiation domination. 
Assuming the primordial power spectrum for $\zeta_{gS}$ in Eq.~\eqref{eq:Lognormal} and fixing $\sigma = 1$, $b (\nu)$ becomes a function of the frequency ratio $\nu / \nu_\ast$, the amplitude $A_S$, as well as the non-Gaussian parameters $\fnl$ and $\gnl$. 
Our numerical results are illustrated in \cref{fig:b_tilde_lin} and \cref{fig:b_all}.

\cref{fig:b_tilde_lin} illustrates the impact of non-Gaussian parameters on $b (\nu)$. 
For example, we compare the profiles of $\fnl = \pm 5$ and $\gnl = \pm 400$ with that of Gaussian $\zeta$. 
The influence of $|\fnl|$ is observed to be more significant than that of $|\gnl|$, while the sign of $\gnl$ has a more distinct impact on the profiles. 
As expected, the non-Gaussianity of \ac{SIGW} background is caused by the non-Gaussian $\zeta$ while disappears for the Gaussian $\zeta$ which is denoted by the black line. 
However, the non-Gaussian $\zeta$ does not always lead to the non-Gaussianity of \ac{SIGW} background across all frequency ranges.
As presented by curves for $\gnl = -400$, which are denoted as the green and red curves, $b (\nu)$ is nearly zero in the frequency band $\nu \sim \nu_\ast$. 
Moreover, for a negative $\gnl$, the angular bispectrum can be negative in lower-frequency ranges if the non-Gaussian parameters satisfy the conditions $(3 \gnl A_S / 5)^2 \lesssim \fnl^2 A_S \lesssim 0.1$ approximately.

\cref{fig:b_all} is an array constituting multiple contour plots with $\fnl$ and $\gnl$ as variables, obtained by selecting different values of $A_S$ and $\nu/\nu_\ast$. 
It is presented to comprehensively elaborate the dependence of the \ac{SIGW} angular bispectrum on $\fnl$ and $\gnl$. 
Each panel is divided into four triangular regions by dotted lines, with the left and right regions representing the contribution from the $\gnl$ term being smaller than that from the $\fnl$ term, and the top and bottom regions representing the contribution from the $\gnl$ term being larger than that from the $\fnl$ term. 
It is worth noting that the value of the angular bispectrum within the ``loop'' in each panel is indeed negative. 
In other words, when $\gnl$ is negative and both $\fnl$ and $|\gnl|$ are relatively mild ($\gnl<0, \fnl^2 A_S \lesssim 0.25, -\gnl A_S \lesssim 0.5$), the angular bispectrum is negative.
Additionally, larger $|\fnl|$ values or negative $\gnl$ values typically result in larger bispectral amplitudes. 
Furthermore, a comparison across different panels reveals that larger values of $A_S$ tend to decrease the reduced bispectral magnitude, whereas a small value of $A_S$ ($10^{-4}$) can lead to the value of the reduced bispectrum even up to $10^{-5}$. 
As the same positions at each panel denote the same values of $\fnl\sqrt{A_S}$ and $\gnl A_S$, the comparison also implies that larger values of $A_S$ or higher frequency ratio $\nu/\nu_\ast$ generally suppress the effects from initial inhomogeneities relatively on the angular bispectrum.

\subsection{Angular trispectrum}\label{sec:num-t}

\begin{figure}[htbp]
    \centering
    \includegraphics[width = 1 \columnwidth]{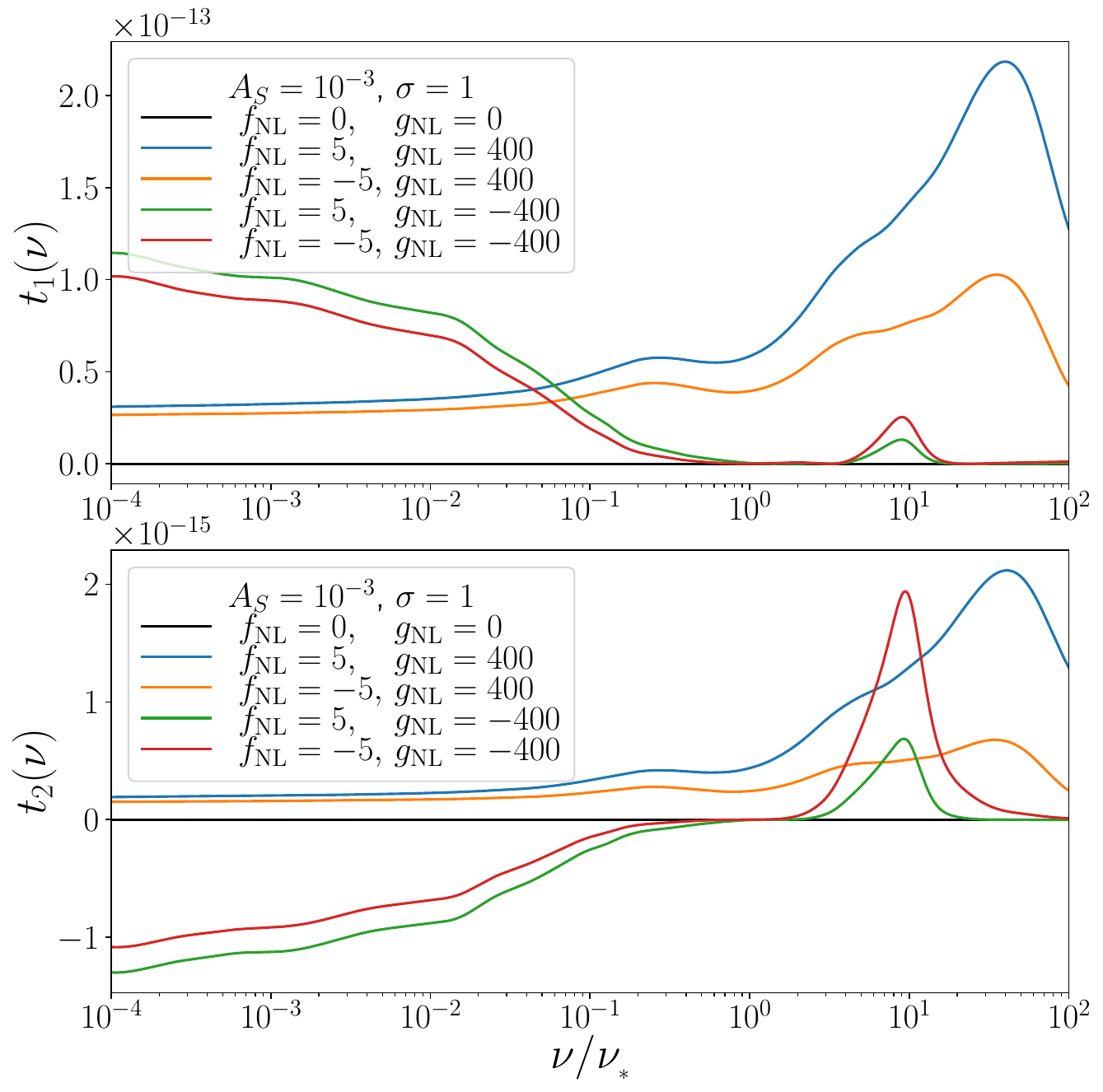}
    \caption{Frequency dependence of the reduced angular trispectrum of SIGWs.  }\label{fig:t_tilde_lin}
\end{figure}

\begin{figure*}[htbp]
    \centering
    \includegraphics[width = 1 \textwidth]{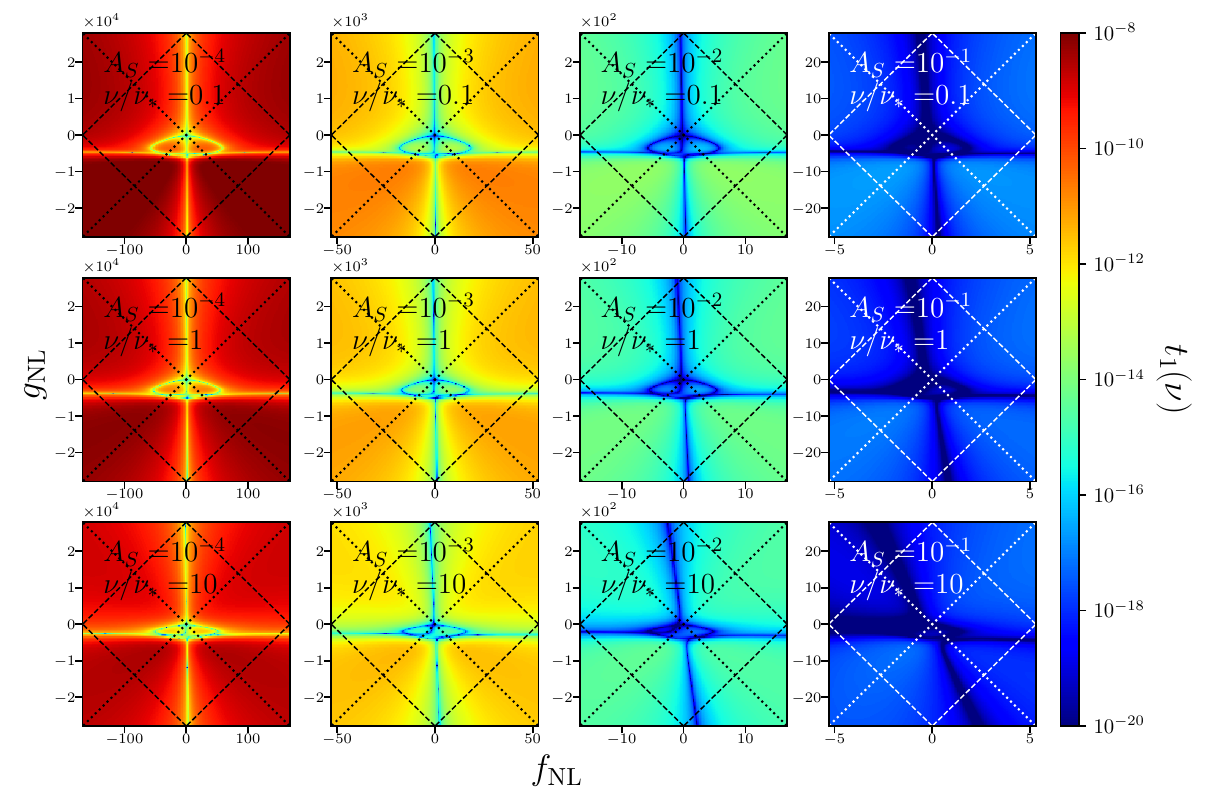}\\
    \includegraphics[width = 1 \textwidth]{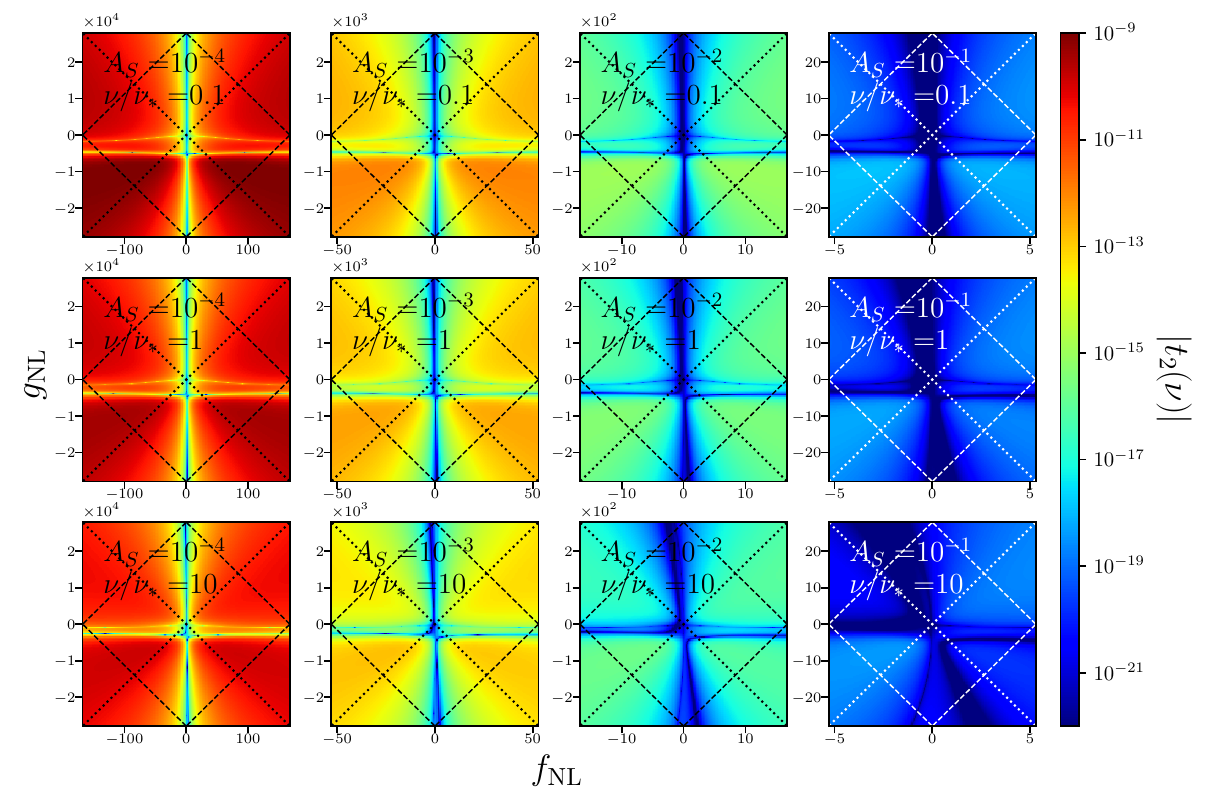}
    \caption{Frequency-dependent functions for the reduced angular trispectrum of SIGWs with respect to the primordial non-Gaussian parameters $\fnl$ and $\gnl$. {  The conventions for the dotted and dashed lines are the same as those in Figure~\ref{fig:b_all}. } 
    }\label{fig:t_all}
\end{figure*}

Following the same step, we examine the two functions, defined as $t_1 (\nu)$ in Eq.~\eqref{eq:t1-def} and $t_2 (\nu)$ in Eq.~\eqref{eq:t2-def}, that characterize the frequency dependence of the reduced angular trispectrum. 
Our numerical results are illustrated in \cref{fig:t_tilde_lin} and \cref{fig:t_all}.

Taking the same values of model parameters as in \cref{fig:b_tilde_lin}, we compare the effects of different values of $\fnl$ and $\gnl$ as well as their signs on the above two functions in \cref{fig:t_tilde_lin}. 
By construction, $t_1 (\nu)$ and $b (\nu)$ behave similarly, apart from that $t_1 (\nu)$ is always nonnegative and $t_1 (\nu) / b (\nu) \sim \zeta_{gL}$. 
The upper panel of \cref{fig:t_tilde_lin} has met our expectation in comparison with \cref{fig:b_tilde_lin}. 
For $t_2 (\nu)$, its magnitude is nearly two orders of magnitude less than that of $t_1 (\nu)$ for all the same sets of model parameters, though both $t_2 (\nu)$ and $t_1 (\nu)$ are of $\cO (\zeta_{gL}^3)$ order. 
This suggests that the contributions from $\langle\delta_{\uGW,0}^{(3)}\delta_{\uGW,0}^{(1)}\delta_{\uGW,0}^{(1)}\delta_{\uGW,0}^{(1)}\rangle$ are subdominant in contrast with that from $\langle\delta_{\uGW,0}^{(2)}\delta_{\uGW,0}^{(2)}\delta_{\uGW,0}^{(1)}\delta_{\uGW,0}^{(1)}\rangle$. 
Moreover, $t_2 (\nu)$ may take a negative value in the low-frequency range if the sign of $\gnl$ is minus, which behaves distinctly from $t_1 (\nu)$ but similarly to $b (\nu)$. 
The profiles of $t_2 (\nu)$ are also influenced by $|\fnl|$ and the sign of $\gnl$ mainly, which are identical to those of $b (\nu)$ and $t_1 (\nu)$. 
In addition, similar to $b (\nu)$, the magnitudes of $t_1 (\nu)$ and $t_2 (\nu)$ are negligible at the frequency band $\nu\sim\nu_\ast$ when $\gnl=-400$.

We also depict two arrays for $t_1 (\nu)$ and $|t_2 (\nu)|$ in \cref{fig:t_all}, which are arranged identically to that of $b (\nu)$. 
We can read the dependence of the \ac{SIGW} angular trispectrum on $\fnl$ and $\gnl$ easily from these arrays. 
While the contour plots of $t_1 (\nu)$ are similar to those of $b (\nu)$ except the magnitudes and signs, the contour plots of $t_2 (\nu)$ possess different features. 
It is observed that $t_2 (\nu) < 0$ for a small value of $|\gnl|$ with a minus sign, regardless the value of $\fnl$. 
Additionally, for both $t_1 (\nu)$ and $t_2 (\nu)$, larger $|\fnl|$ values or negative $\gnl$ values tend to increase their magnitudes while larger values of $A_S$ always suppress them, which are same as that of $b (\nu)$. 
It is telling that the non-Gaussianity of \ac{SIGW} background is more prominent in these cases. 
Though the magnitude of $t_2 (\nu)$ is typically one order of magnitude smaller than that of $t_1 (\nu)$, their different dependence on frequency and multipoles implies that the reduced angular trispectrum contains a wealth of information about the early universe.

\section{Discussion and Conclusion}\label{sec:conclusion}

In this study, we have delved into the non-Gaussianity of \ac{SIGW} background associated with the local-type primordial non-Gaussianity parameterized by $\fnl$ and $\gnl$. 
For the first time, we have derived the explicit formulae for the angular bispectrum and trispectrum of \acp{SIGW}, which are complementary to our existing work \cite{Li:2023xtl} that devoted to the study of the energy-density fraction spectrum and angular power spectrum. 

Based on the diagrammatic approach used in Ref.~\cite{Li:2023xtl}, we have generalized the computational technique to fully calculate the \ac{SIGW} density contrast up to the $\cO (A_L^3)$ order, and derived the comprehensive expressions for the reduced angular bispectrum $\tilde{b}_{\ell_1 \ell_2 \ell_3} (\nu)$ and trispectrum $t^{\ell_1 \ell_2}_{\ell_3 \ell_4} (L,\nu)$. 
Notably, these spectra were found to depend on frequency bands in a non-trivial way, as their information is unable to be compressed into the energy-density fraction spectrum $\Omega_{\uGW,0} (\nu)$ and the reduced angular power spectrum $\tilde{C}_\ell (\nu)$. 
This suggests that the non-Gaussianity of \ac{SIGW} background contains additional information about the \ac{SIGW} background beyond the energy-density fraction spectrum and the angular power spectrum. 
Furthermore, the expressions for $\tilde{b}_{\ell_1 \ell_2 \ell_3} (\nu)$ and $t^{\ell_1 \ell_2}_{\ell_3 \ell_4} (L,\nu)$ have verified our prior expectation that the non-Gaussianity of \ac{SIGW} background results from the primordial non-Gaussianity of cosmological curvature perturbations. 
Only when different modes of primordial curvature perturbations are coherent, \acp{SIGW} at two distant locations are coherent. 
We also found that for a negative value of $\gnl$, a positive \ac{SIGW} angular bispectrum indicates significant primordial non-Gaussianity. 
Eventually, our numerical results reveal the substantial impact of primordial non-Gaussianity on the angular bispectrum and trispectrum, with the bispectral amplitude reaching $\sim10^{-5}$ and the trispectral amplitude reaching $\sim10^{-8}$ for specific sets of model parameters.

The future advancements in the angular resolution of \ac{GW} detectors will allow possible extractions of valuable information from the detection of the stochastic \ac{GW} background, which can be produced from the superposition of various sources.
Recently, multiple \ac{PTA} experiments \cite{Xu:2023wog,EPTA:2023fyk,NANOGrav:2023gor,Reardon:2023gzh} have announced the existence of a stochastic \ac{GW} background with compelling evidence. 
However, the datasets did not affirm a certain \ac{GW} source for the signal, with various possibilities discussed by accompanying papers \cite{Antoniadis:2023xlr,NANOGrav:2023hvm}. 
Subsequent studies \cite{Franciolini:2023pbf,Inomata:2023zup,Cai:2023dls,Wang:2023ost,Liu:2023ymk,Abe:2023yrw,Ebadi:2023xhq,Figueroa:2023zhu,Yi:2023mbm,Madge:2023cak,Firouzjahi:2023lzg,Zhu:2023faa,You:2023rmn,Ye:2023xyr,HosseiniMansoori:2023mqh,Balaji:2023ehk,Das:2023nmm,Bian:2023dnv,Jin:2023wri,Zhao:2023joc,Liu:2023pau,Yi:2023tdk,Frosina:2023nxu,Choudhury:2023hfm,Ellis:2023oxs,Kawasaki:2023rfx,Yi:2023npi,Harigaya:2023pmw,An:2023jxf,Gangopadhyay:2023qjr,Chang:2023ist,Inomata:2023drn,Choudhury:2023fwk,Choudhury:2023fjs,Domenech:2023jve,Chang:2023aba,Mu:2023wdt,Choudhury:2024one,Chen:2024twp} speculated that this signal potentially originates from \acp{SIGW}. 
The authors provided an avenue for testing and validating the presence of \acp{SIGW} and the associated non-Gaussian features. 
In particular, Refs.~\cite{Wang:2023ost,Li:2023xtl} showed that if the \ac{PTA} signal originates from \acp{SIGW}, it can be tested with the \ac{SKA} \cite{2009IEEEP..97.1482D,Weltman:2018zrl,Moore:2014lga} by measuring the corresponding angular power spectrum. 
If the angular resolution of \ac{SKA} is sufficient enough to observe the anisotropies in the stochastic \ac{GW} background, there is a promising opportunity to measure its angular bispectrum and trispectrum, besides the energy-density fraction spectrum and the angular power spectrum, thereby further confirming the sources of the signal and even determining the relevant model parameters. 

Our present study would be helpful in search of \acp{PBH}, which are hypothesized to be a reasonable candidate of cold dark matter and a potential origin of the individual \ac{GW} events observed by the Advanced \acl{LVK} \cite{LIGOScientific:2018mvr,LIGOScientific:2020ibl,LIGOScientific:2021djp,Wang:2022nml,Domenech:2024cjn}. 
As emphasized in the prior works 
\cite{Bullock:1996at,Byrnes:2012yx,Young:2013oia,Franciolini:2018vbk,Passaglia:2018ixg,Atal:2018neu,Atal:2019cdz,Taoso:2021uvl,Meng:2022ixx,Chen:2023lou,Kawaguchi:2023mgk,Fu:2020lob,Inomata:2020xad,Young:2014ana,Choudhury:2023kdb,Garcia-Bellido:2017aan,Nakama:2016gzw,Ferrante:2022mui,Green:2020jor,Carr:2020gox,Escriva:2022duf,Escriva:2022pnz,Ezquiaga:2019ftu,Kehagias:2019eil}, the mass function of \acp{PBH} is significantly influenced by the presence of primordial non-Gaussianity, as their formation threshold can be affected by specific characteristics of the non-Gaussianity. 
It is worth noting that the angular bispectrum and trispectrum of \ac{SIGW} background are also significantly affected by the primordial power spectrum. 
Therefore, future measurements of the non-Gaussianity of this \ac{SIGW} background may further exert constraints on the \ac{PBH} scenarios. 
In addition, the couplings between short- and long-wavelength modes also cause an inhomogeneous distribution of \acp{PBH} on large scales \cite{Papanikolaou:2024kjb}. 
If \acp{PBH} contribute to dark matter, the primordial non-Gaussianity produces the isocurvature perturbations of dark matter during radiation domination \cite{Franciolini:2018vbk,Young:2015kda}. 
The \ac{CMB} constraints on these isocurvature modes from the \texttt{Planck} satellite \cite{Planck:2018jri} could impose limits on the local-type primordial non-Gaussianity \cite{Tada:2015noa,Bartolo:2019zvb}. 
They are expected to be further enhanced by future observations of \ac{CMB} and \ac{LSS}. 
We would not delve into a detailed study of them in this work, but defer it to future works. 

In this work, we have focused on local-type primordial non-Gaussianity with scale-invariance, but our approach can be extended to more general cases. 
For clarification, the energy-density fraction spectrum is sensitive to the primordial non-Gaussianity on small scales, while the anisotropy is related with the primordial non-Gaussianity that characterize couplings between short- and long-wavelength curvature perturbations, although we treat them as the same in this work.  
If $\fnl$ and $\gnl$ change gradually across frequency bands, we can treat them as constants when evaluating the energy-density fraction spectrum and simply use varying values in the numerators in Eqs.~(\ref{eq:Ong1-def},\ref{eq:Ong2-def},\ref{eq:Ong3-def}), similar to the method used in Ref.~\cite{Yu:2023jrs}. 
Otherwise, if the scale-dependence of $\fnl$ and $\gnl$ is non-negligible in the relevant frequency bands, we would need to incorporate them into the integrands for the components of energy-density fraction spectrum and integrate these integrands as a whole, which would significantly complicate the numerical calculation though the semi-analytical expressions are similar. 
It is worth noting that this method is also available beyond the postulation of local-type, which has been adopted in Ref.~\cite{Ragavendra:2021qdu}. 
Moreover, other shapes of primordial non-Gaussianity, rather than the local-type considered in our present work, are unlikely to cause significant anisotropy and non-Gaussianity of \ac{SIGW} background, as the deviations from isotropic and Gaussian \ac{SIGW} background stem from the couplings between short- and long-wavelength modes.

In conclusion, our work is of great significance for future detection of stochastic \ac{GW} background and exploration of the early universe and dark matter. 
Measurements of angular bispectrum and trispectrum of stochastic \ac{GW} background would provide a vital complement to probe the primordial non-Gaussianity beyond the sensitive regimes of \ac{CMB} and \ac{LSS}. 
We would like to highlight that our research approach can be easily extended to study the primordial non-Gaussianity of higher orders than $\gnl$ or with the scale-dependence.


\begin{acknowledgments}
We appreciate Mr. Yan-Heng Yu for helpful discussion. 
S.W. and J.P.L. are partially supported by the National Natural Science Foundation of China (Grant No. 12175243), the National Key R\&D Program of China No. 2023YFC2206403, the Science Research Grants from the China Manned Space Project with No. CMS-CSST-2021-B01, and the Key Research Program of the Chinese Academy of Sciences (Grant No. XDPB15).  Z.C.Z. is supported by the National Key Research and Development Program of China Grant No. 2021YFC2203001 and the National Natural Science Foundation of China (Grant NO. 12005016). K.K. is supported by KAKENHI Grant No. JP22H05270. 

\end{acknowledgments}

\bibliography{biblio}
\bibliographystyle{JHEP}

\end{document}